\documentclass[11pt]{article}
\usepackage{fullpage}
\usepackage{times}
\usepackage{cite}
\usepackage{latexsym}
\usepackage{amsmath}
\usepackage{amssymb} % for \mathbb
\usepackage{graphics}
\usepackage{color}

\graphicspath{{.}}

\newtheorem{theorem}{Theorem}[section]
\newtheorem{lemma}[theorem]{Lemma}
\newtheorem{corollary}[theorem]{Corollary}

\newtheorem{fact}[theorem]{Fact}
\newtheorem{definition}[theorem]{Definition}

\newcommand{\qed}{\hfill $\Box$}
\newenvironment{proof}{{\bf Proof.~}}{\qed}

\newcommand{\ignore}[1]{}

\newcommand{\dg}{\text{\it deg}}
\newcommand{\desc}{\text{\it desc}}
\newcommand{\depth}{\text{\it depth}}
\newcommand{\poly}{\text{\rm poly}}
\newcommand{\bmax}{B^*}

\newcommand{\CA}{\ensuremath{{\cal A}}}
\newcommand{\CE}{\ensuremath{{\cal E}}}
\newcommand{\CG}{\ensuremath{{\cal G}}}
\newcommand{\II}{\ensuremath{{\cal I}}}

\newcommand{\CR}{\ensuremath{{\cal R}}}

\newcommand{\fig}[3]{%
	\begin{figure}[t]%
	\centerline{\includegraphics{#1}}%
	\caption{#2}%
	\label{#3}%
	\end{figure}%
}

\begin{document}

\title{Rectangular Layouts and Contact Graphs}

\author{Adam L.~Buchsbaum\thanks{
		AT\&T Labs,
		Shannon Laboratory,
		180 Park Ave.,
		Florham Park, NJ 07932,
		\{alb,erg,magda,suresh\}@research.att.com.
	}
   \and Emden R.~Gansner$^*$
   \and Cecilia M.~Procopiuc$^*$
   \and Suresh Venkatasubramanian$^*$
}

\date{October 11, 2006}

\maketitle

\begin{abstract}
Contact graphs of isothetic rectangles
unify many concepts
from applications including VLSI and architectural design,
computational geometry, and GIS.
Minimizing the area of their corresponding {\em rectangular layouts}
is a key problem.
We study the area-optimization problem
and show that it is NP-hard to find a
minimum-area rectangular layout
of a given contact graph.
We present $O(n)$-time algorithms
that construct $O(n^2)$-area rectangular layouts
for general contact graphs
and $O(n\log n)$-area rectangular layouts for trees.
(For trees, this is an $O(\log n)$-approximation algorithm.)
We also present an infinite family of graphs (rsp., trees)
that require $\Omega(n^2)$ (rsp., $\Omega(n\log n)$) area.

We derive these results
by presenting a new characterization
of graphs that admit
rectangular layouts
using the related concept of {\em rectangular duals}.
A corollary to our results
relates the class of graphs that admit rectangular layouts
to {\em rectangle of influence drawings}.

\end{abstract}

\section{Introduction}
\label{sec:intro}

Given a set of objects in some space,
the associated {\em contact graph}
contains a vertex for each object
and an edge implied by each pair of objects
that touch in some prescribed fashion.
While contact graphs have been extensively studied
for objects such as curves, line segments, and even strings 
	(surveyed by Hlin\v{e}n\'{y} \cite{ccg:h98} and
	Hlin\v{e}n\'{y} and Kratochv\'{\i}l \cite{contact:hk01}),
as has the more general class of {\em intersection graphs}
	(surveyed by
	Brandst\"{a}dt, Le, and Spinrad \cite{gc:bls99}
	and McKee and McMorris \cite{ig:mm99}),
the literature on {\bf closed} shapes is relatively sparse.
Koebe's Theorem \cite{koebe:k36}
states that any 
planar graph (and only a planar graph)
can be expressed as the contact graph of disks in the 
plane.\footnote{
	This result was lost and recently rediscovered 
	independently by Andreev and Thurston;
	Sachs \cite{contact:s94} provides a history.
}
More recently, de Fraysseix, Ossona de Mendez and Rosenstiehl
\cite{tri:for94}
show that any planar graph can be represented
as a triangle contact graph but not vice-versa.
In this paper,
we consider contact graphs of isothetic rectangles.

Contact graphs of rectangles find critical applications
in areas including
VLSI design \cite{fp:ll88,slice:sb88,slice:sb91,fp:ttss91,slice:ys95},
architectural design \cite{arch:s76},
and, in other formulations,
computational geometry \cite{dom:bco92,rvis:ow88}
and
geographic information systems \cite{rdgis:gs69}.
Previous works
considered concepts similar to contact graphs
but used a variety of notions
like rectangular duality \cite{rdual:h93,rdual:kk85,rdual:kk88,rd:ll90}
and concepts from graph drawing
such as orthogonal, 
rectilinear, 
visibility, and 
proximity layouts \cite{graphdraw:bett99,prox:dll94,rpl:rt86}
as well as rectangular drawings \cite{brd:h01,brd:rnn00,rd:rng04,rgd:rnn98}
to achieve their results.
Our work is the first to deal directly with
contact graphs of rectangles,
yielding a simpler foundation for study.

We call
a collection of rectangles that realizes a given contact graph $G$
a {\em rectangular layout}:
a set of disjoint, isothetic rectangles
corresponding to the vertices of $G$,
such that two rectangles are adjacent
if and only if the implied edge exists in $G$.
The associated optimization problem
is to find a rectangular layout minimizing
some criterion such as area, width, or height.
This problem is inherently intriguing, with many enticing subproblems and
variations.  
It arises in practice in the design of an interface to a relational
database system used in AT\&T to allow customers to model and administer
the equipment and accounts in their telecommunications networks.  This
system allows users to specify their own schemas using entity-relationship
models \cite{dbms:rg00}.  The system then presents the database entities as
rectangular buttons.  Clicking on a button provides information related to
the corresponding entity type:  e.g., detailed descriptions of the entity
attributes or information about specific records.  Experience indicates
the benefit of juxtaposing buttons that correspond to related entities.
Viewing the database schema as the obvious graph, with entities
as vertices and relations as edges, leads to rectangular layouts.
Solving the related optimization problems would automate this part
of tailoring the interface to the schema, of significant benefit as
otherwise each customer's interface must be built manually.

We solve several important problems
concerning rectangular layouts.
We give a new characterization of planar graphs that admit 
rectangular layouts
in terms of those that can be embedded without {\em filled triangles}.
En route, we unify a number of different lines of research in this field.
We show a suite of results concerning the hardness
of finding optimal layouts;
design algorithms to construct layouts on graphs, 
and, with better area bounds, trees;
and present some worst-case area lower bounds.
We detail these results at the end of this section.

\subsection{Relationship to Prior Work}

Rectangular layouts are dual concepts to {\em rectangle drawings}
of planar graphs:  that is,
straight-line, isothetic embeddings 
with only rectangular faces.
Recent work on rectangular drawings \cite{rd:rng04,rgd:rnn98}
and the related box-rectangular drawings \cite{brd:h01,brd:rnn00}
culminates with Rahman, Nishizeki, and Ghosh's linear-time
algorithm for finding a rectangular drawing of any planar graph
if one exists \cite{rd:rng04}.
In a rectangular layout, the rectangles themselves
correspond to vertices.
While the two concepts are more or less dual to each other,
moving between them can be highly technical.
In particular, the machinery to find rectangular drawings
of graphs that
are not three-connected is complex.
Our contribution gives a direct method for constructing
rectangular layouts, and we handle cases of low connectivity
easily.

Rectangular layouts are closely related
to {\em rectangular duals},
which are like layouts except
that a rectangular dual must form a dissection of
its enclosing rectangle;
i.e., it allows no gaps between rectangles.
Rectangular duals have a rich history,
including much work on 
characterizing graphs that admit rectangular duals \cite{rdual:h93,rdual:kk85,rdual:kk88,rd:ll90},
transforming those that do not by adding new vertices \cite{st:aav00,fp:ll88},
and constructing rectangular duals in linear time \cite{rd:bs88,rdual:h93,rel:kh97}.
The proscription of gaps, however,
severely limits the class of graphs that admit rectangular duals;
for example, paths are the only trees that have such duals.
In general, any (necessarily planar) graph admitting a rectangular dual
must be internally triangulated,
but no such restriction applies to layouts.
This simple distinction
yields many advantages to layouts over duals.
The cleaner, less specified definition of layouts
characterizes a class of graphs
that is both more general (including all trees, for example)
and also much simpler to formalize.
When we discuss area,
we will also show that while different variations
of rectangular layouts
have different degrees of area monotonicity under
graph augmentation,
rectangular duals do not enjoy any such monotonicity:
a small graph might require a significantly larger dual 
than a larger graph.
Additionally, there are graphs that admit asymptotically
smaller rectangular layouts than duals.

Another closely related area concerns VLSI {\em floorplanning},
in which an initial configuration of rectangles (usually
a dissection) is given,
the goal being to rearrange and resize the rectangles
to minimize the area
while preserving some properties
of the original layout.
While there is an extensive literature on floorplanning,
there is much divergence within it as to what
criteria must be preserved during minimization.
These typically include minimum size constraints on the rectangles
plus one of various notions of adjacency equivalence.
Stockmeyer \cite{slice:s83} presents one of the cleanest
definitions of equivalence,
preserving the notion of relative placement
of rectangles (whether one appears
left of, right of, above, or below another);
other criteria include the preservation of 
relative area and aspect ratios \cite{fp:wkc88}
and that of relative lengths of abutment \cite{fp:ttss91}.
These works are further classified by {\em sliceability}:
a floorplan is {\em sliceable}
if it can be recursively deconstructed
by vertical and horizontal lines extending
fully across the bounding box.
Minimizing the area of non-sliceable floorplans
has been shown to be NP-hard under various
constraints \cite{thesis:aslp80,fp:psl96,slice:s83},
whereas area minimization of sliceable floorplans
is tractable \cite{fp:s96,slice:s83,fp:ys93}.
Not all floorplans can be realized 
by sliceable equivalents \cite{slice:sb88,slice:sb91},
so work exists on 
isolating and generating sliceable floorplans where possible \cite{slice:ys95}
as well as minimizing the area of non-sliceable floorplans
by various heuristics \cite{fp:psl96,fp:ttss91,fp:wkc88}.

Two major facets distinguish floorplanning from 
area-minimization of rectangular layouts.
First, floorplanning seeks to minimize the area
of an arrangement of rectangles
given a priori by some external process (possibly human design);
rectangular layouts themselves are determined by corresponding contact graphs.
Second, the notion of equivalence among floorplans
differs from context to context,
whereas contact graph adjacencies strictly
identify the equivalence of rectangular layouts.
Thus,
there can exist different rectangular layouts
of the same graph that do not represent equivalent
floorplans,
even
by Stockmeyer's definition;
and conversely
there can exist equivalent floorplans
that are layouts for non-isomorphic graphs.
Still, much work in floorplanning
uses concepts from rectangular duals,
so work on rectangular layouts
can also contribute to this area.

Finally, also related is the idea of
{\em proximity drawings} \cite{prox:dll94,rng:jt92},
in which a set of objects corresponds to the vertices of a graph,
with edges connecting vertices of correspondingly close objects
for some definition of proximity.
A particularly relevant special case 
is that of {\em rectangle of influence drawings} \cite{rid:llmw98},
which are (not necessarily planar) straight-line embeddings
of graphs
such that the isothetic rectangles induced by pairs of vertices
contain no other vertices if and only if the corresponding edges exist.
Using results of Biedl, Bretscher, and Meijer \cite{rid:bbm99},
we show that graphs that admit rectangular layouts
are precisely those that admit a weaker variation
of planar rectangle of influence drawings,
in which induced rectangles may be empty even if
the corresponding edges are missing from the graph;
i.e., that contact graphs of rectangles
also express this variation of rectangle of influence drawings.

\subsection{Our Results}

The many parallel lines of research (in different communities)  in the
general area of contact graphs of rectangles have led to overlapping (and
in some cases equivalent) definitions and results. 
Our contributions
in this paper are two-fold: we present algorithms for optimizing
rectangular layouts and prove various hardness results, 
and we also prove
various structural results that tie existing work together in a
coherent way
that produces efficient algorithms.

\begin{itemize}
\item We provide a new characterization
	of the class of graphs that admit rectangular
  layouts. This characterization is equivalent to an earlier result by
  Thomassen \cite{rl:t86} and has the added advantage of yielding an $O(n)$-time
  algorithm for checking if a given $n$-vertex planar graph 
	admits such a layout.
	Moreover, we can construct a layout for the input graph in
  $O(n)$ time. We also prove an upper bound of $O(n^2)$ on the area of the layout,
  and we give a matching worst-case lower bound.
\item We give an $O(n)$-time algorithm that constructs an $O(n\log n)$-area
	layout for any tree.
  We also demonstrate a general class of trees that
  are \emph{flexible}: i.e., they can be laid out in linear area with
  \emph{any aspect ratio}. 
  Finally we show that in general trees cannot have arbitrarily thin
	layouts:  
  there exists a class of trees such that the minimum dimension must
   have size $\Omega(\log n)$.
This bound uses solely topological
  arguments and may be of independent interest. 
  In particular, it leads to an $\Omega(n\log n)$ worst-case
	area lower bound, matching the upper bound of our algorithm.
\item 
  We prove that the problem of optimizing the area of a rectangular
  layout is NP-hard.  The proof also shows that optimizing the width given a
  fixed height, or vice-versa, is NP-hard.
\item 
  We 
  %also 
  show that rectangular duals can be much larger than layouts;
  there exists a class of graphs having $O(n)$-area layouts but
  $\Omega(n^2)$-area duals. 
\item Our characterization of contact graphs of rectangles in terms of
\emph{filled triangles} establishes a connection between
rectangular layouts and \emph{rectangle of influence drawings}. 
Specifically, a corollary of our characterization is that the
class of graphs having rectangular layouts is identical to the class of
graphs having planar, weak, closed rectangle of influence drawings\cite{rid:bbm99}.
\end{itemize}

\paragraph{Paper Outline.}

We continue in Section \ref{sec:defn} by introducing necessary definitions,
	including those for rectangular layouts themselves,
	and we also state some useful lemmas.
In Section \ref{sec:layouts}
	we present our new characterization of graphs
	that admit rectangular layouts.
We use this characterization in Section \ref{sec:graphs}
	to design our linear-time algorithm for
	constructing $O(n^2)$-area rectangular layouts,
	and we show matching worst-case lower bounds.
In Section \ref{sec:trees}
	we present improved results for trees.
We present hardness results in Section \ref{sec:area},
and we conclude in Section \ref{sec:conc}.

\section{Definitions}
\label{sec:defn}

Throughout, we assume without loss of
generality that all graphs are connected
and have at least 4 vertices.
For a given graph $G=(V,E)$, define
$n=|V|$ and $m=|E|$.
We say $G'=(V',E')$ is a {\em subgraph} of $G$
if $V'\subseteq V$ and $E' \subseteq E$;
$G'$ is a {\em proper subgraph} of $G$
if $G'$ is a subgraph of $G$ and $G'\not=G$.

A graph $G$ is \emph{$k$-connected} if the removal
of any set of $k-1$ vertices
leaves the remainder of $G$ connected.
A \emph{separating triangle} of $G$ is a 3-vertex cycle
whose removal disconnects the remainder of $G$.  For an arbitrary but
fixed embedding of a planar graph, a {\em filled triangle} is defined to
be a length-3 cycle with at least one vertex 
inside the induced region. 
Note that an embedding of a graph might have a
filled triangle but no separating triangle
(e.g., any embedding of $K_4$)
and vice-versa.
A \emph{cubic graph} is a
regular degree-3 graph.  A {\em planar triangulation} is a planar graph in
which all faces 
are bounded by 3-vertex cycles.  Whitney's 2-Isomorphism Theorem
\cite{g:w33} 
in fact implies that a planar triangulation has a unique embedding up
to stereographic projection, which preserves the facial structure, so it
is well-defined to talk about a planar graph $G$ itself being a
triangulation (or {\em triangulated}) as opposed to some specific
embedding of $G$ being triangulated.  Note also that a planar
triangulation $G$ has a planar dual, denoted $G^*$, which is unique up to
isomorphism.  Clearly, any 4-connected graph has no separating triangle;
for planar triangulations, the converse holds as well.

\begin{lemma}
\label{lem:4ctri}
A planar triangulation $G$ is 4-connected
if and only if $G$ has no separating triangles.
\end{lemma}

Ko\'{z}mi\'{n}ski and Kinnen
\cite[Lem.~1]{rdual:kk85}
state Lemma \ref{lem:4ctri} in terms of a fixed embedding,
but the above remarks obviate the issue of fixed embeddings
for planar triangulations.

A graph $G$ is {\em cyclically $k$-edge
  connected} if the removal of any $k-1$ edges either leaves $G$ connected
or else produces at least one connected component that contains no cycle,
i.e., does not break $G$ into multiple connected components all of which
contain cycles \cite{gt:w96}. Cyclic 4-edge connectivity is a dual concept to
4-connectivity for planar triangulations. 
In particular, we will use the following result,
which appears to be well known;
we prove it in Appendix \ref{app:proofs} for completeness.

\newcommand{\lemptc}{
A graph $G$ is a 4-connected planar triangulation
if and only if $G^*$ is planar, cubic, and cyclically 4-edge connected.
}
\begin{lemma}
\label{lem:ptc4ec}
\lemptc
\end{lemma}

Define a {\em rectangular dissection} to be a partition of a rectangle
into smaller rectangles, no four of which meet in any point.  A {\em
  rectangular representation} of a graph $G$ is a rectangular dissection
$R$ such that the vertices of $G$ map 1-to-1 to the intersections of line
segments of $R$ minus the four external corners of $R$; i.e., $R$ is a
straight-line, isothetic drawing of $G$ except for the four edges that
form right angles at the external corners of $R$.  A {\em rectangular
  dual} of a graph $G$ is a rectangular dissection, if one exists, whose
geometric dual minus the exterior vertex and incident edges is $G$.

Rectangular duals allow no {\em gaps}; i.e., areas within
the bounding box that do not correspond to any vertices.
We introduce the concept of {\em rectangular layout}
to allow gaps in the representation.
This allows a larger class of graphs to be represented.

\begin{definition}[Rectangular Layout]
A {\em (strong) rectangular layout} of a graph $G=(V,E)$
is a set $R$ of isothetic rectangles
whose interiors are pairwise disjoint,
with
an isomorphism $\CR : V \rightarrow R$
such that for any two vertices $u,v \in V$,
the boundaries of $\CR(u)$ and $\CR(v)$ overlap non-trivially
if and only if $\{u,v\}\in E$.
\end{definition}

See Fig.~\ref{fig:example}.
\fig{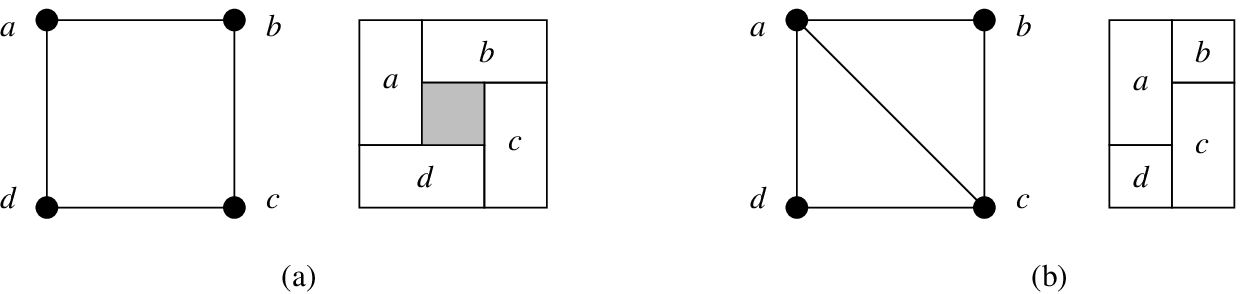}{Two graphs and associated rectangular layouts.
        \ignore{In this and succeeding depictions of layouts, shaded}
        Shaded 
        regions depict gaps.}{fig:example}
The requirement that {\bf non-trivial} boundary overlaps
define adjacencies (that rectangles meeting
at corners are not considered adjacent)
is significant:
allowing corner-touching to imply adjacency
changes the class of graphs expressed; cf., Section \ref{sec:conc}.
This therefore obviates the specification that the rectangles be isothetic:
any collection of rectangles inducing a {\bf connected} contact graph
must be isothetic.
In contrast, there is no a priori proscription on trivial corner touching
in layouts themselves.
We shall show, however,
that we can exclude such trivial corner touching without loss of generality.

We also define a relaxed variation:
A {\em weak rectangular layout} of $G$
is a set $R$ of isothetic rectangles
whose interiors are pairwise disjoint,
with
an isomorphism $\CR : V \rightarrow R$
such that for any edge $\{u,v\} \in E$,
the boundaries of $\CR(u)$ and $\CR(v)$ overlap non-trivially.
In a weak layout, two rectangle boundaries may overlap
even if the corresponding vertices are not adjacent.
For example, in Fig.~\ref{fig:example},
layout (b) is a weak layout for both graphs (a) and (b)
but a strong layout only for graph (b).

Consider a layout drawn on the integral grid.
The {\em area} of the layout is 
that of the smallest enclosing isothetic bounding box.
While corresponding weak and strong layouts may have different areas,
from a feasibility standpoint the distinction does not matter,
as shown by the following lemma.
(Clearly any strong layout is also a weak layout.)

\begin{lemma}
\label{lem:weak}
If $G$ has a weak rectangular layout $R$,
then $G$ has some strong rectangular layout $L$.
\end{lemma}
\begin{proof}
Consider any two rectangles $a$ and $b$ in $R$
whose boundaries overlap but such that
$\{a,b\}\not\in E$.
If the overlapping boundary of one (say $a$)
is completely nested in that of the other ($b$),
we can separate the two rectangles
by moving the touching boundary of $a$
away from $b$ by some small amount,
$\epsilon$,
that does not affect any other rectangle adjacency.
Formally,
let $\epsilon'$ be the size of the smallest boundary overlap in $R$;
then let $\epsilon = \epsilon'/2$.
(On the integral grid, first scale all coordinates by a factor of 2,
and then $\epsilon=1$.)
Assume $a$'s bottom boundary nests within $b$'s top boundary;
other cases are symmetric.
Move $a$'s bottom boundary up by $\epsilon$,
leaving its other boundaries untouched; that is,
$a$ is shrunk, not translated.
See Figure \ref{fig:w2s}(a).
	\begin{figure}[t]%
	\begin{center}
	\input{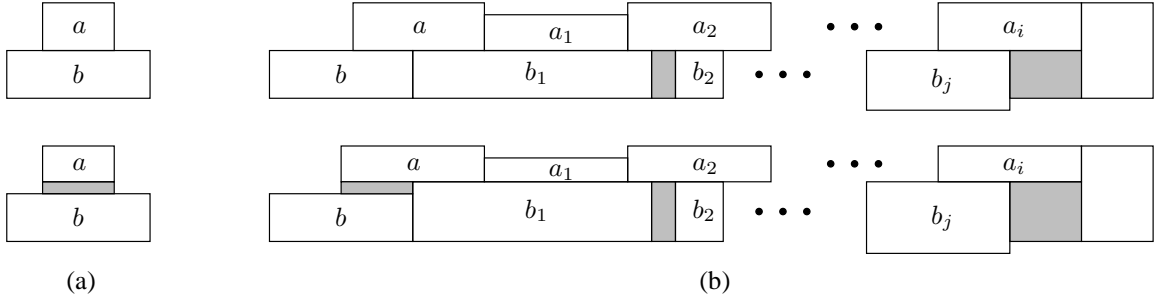}
	\end{center}
	\caption{Weak layouts (top) and corresponding strong layouts
	(bottom) after fixing the $\{a,b\}$ violation.
	(a) The violating boundary of $a$ is nested in that of $b$.
	(b) No nesting occurs; shear the projection of the
	violating boundary until hitting a blocking rectangle.
	In both cases, there can be other rectangles not shown,
	but if the new separation $\epsilon$ is less than the smallest
	prior boundary overlap, no other adjacencies are broken.}%
	\label{fig:w2s}%
	\end{figure}%

By construction of $\epsilon$, any rectangle touching $a$
on either its left or right side still touches $a$,
and $a$'s top boundary remains unchanged, so
the only rectangle adjacency that is broken is that
between $a$ and $b$.

If neither violating boundary is nested in the other,
assume $a$ is to the right and top of $b$; other cases are symmetric.
Consider the projection of the violating boundary rightwards
until it hits some other rectangle or the right side of the bounding box.
In left-to-right order,
call the rectangles with bottom boundaries along this projection
$a_1,a_2,\ldots,a_i$;
and
call the rectangles with top boundaries along this projection
$b_1,b_2,\ldots,b_j$.
See Figure \ref{fig:w2s}(b).
(Note there may be gaps along this projection.)
Move the bottom boundaries of $a,a_1,a_2,\ldots,a_i$
and the top boundaries of $b_1,b_2,\ldots,b_j$
up by $\epsilon$
(as defined above).
This fixes the $\{a,b\}$ violation.
Furthermore, any rectangle adjacent to the left side of $a$
remains so by definition of $\epsilon$.
Other boundary relationships among the adjusted rectangles
are preserved, and no other boundaries are affected.

Repeating this process to fix each violation produces $L$.
\end{proof}

Lemma \ref{lem:weak} addresses feasibility of rectangle layouts
only; the expansion in area from the transformation might be
exponential.  Later we give procedures to draw strong layouts
directly with better areas.

In the sequel, we
will use the term \emph{layout} to refer to strong layouts.
Furthermore, the assumption that no two rectangles meet
trivially at a corner is without loss of generality.
Say rectangles $a$ and $b$ so meet.
We can perturb the boundaries by some small amount
to make the boundary overlap non-trivial.
The layout becomes weak if it was not already.
Lemma \ref{lem:weak} shows that it can be made strong
with only non-trivial boundary overlaps.

\section{Characterizing Rectangular Layouts}
\label{sec:layouts}

\subsection{Background}
\label{ssec:bkgrnd}

Thomassen characterizes graphs that admit rectangular layouts (which he
calls {\em strict rectangle graphs}) as precisely the class of proper
subgraphs of 4-connected planar triangulations \cite[Thm.~2.1]{rl:t86}.
Together with earlier work \cite{rl:t84} his work yields a polynomial-time
algorithm for testing a graph $G$ to see if it admits a rectangular layout
and, if so, constructing such a layout.  He does not analyze the algorithm
precisely for running time,
however,
nor does he bound the layout area at all,
two criteria that concern us.

Thomassen's work rests critically on earlier work by Ungar~\cite{rd:u53}.
Ungar defines a {\em saturated plane map}
to be a finite set of non-overlapping regions 
that partitions the plane
and satisfies the following conditions.
\begin{enumerate}
\item Precisely one region is infinite.
\item At most three regions meet in any point.
\item\label{spm:iii} Every region is simply connected.
\item The union of any two adjacent regions is simply connected.
\item\label{spm:v} The intersection of any two regions
        is either a simple arc or is empty.
\end{enumerate}

An {\em $n$-ring} is a set of $n$ regions
such that their union
is multiply connected.\footnote{
	{\em Connectivity} in this definition and conditions
	\ref{spm:iii}--\ref{spm:v} above is in the topological sense.}
Ungar \cite[Thm.~A]{rd:u53} shows
that a saturated plane map 
that contains no 3-ring is isomorphic
to a rectangular dissection.
Ungar's Theorem B \cite{rd:u53}
further implies that for any rectangular dissection $R$
and any two adjacent rectangles $a$ and $b$ in $R$,
there exists a rectangular dissection $R'$
isomorphic to $R$
such that $a'$ (the region in $R'$ corresponding to $a$ in $R$)
is the infinite region
and $b'$ (the rectangle in $R'$ corresponding to $b$ in $R$)
has three or four whole sides {\em fully exposed}:
i.e., not overlapping any rectangle other than $a'$.

We prove the following ``folklore'' lemma in 
Appendix~\ref{app:proofs}. 
\newcommand{\lemcc}{
A graph $G$ is planar, cubic, and cyclically 4-edge connected
if and only if
$G$ is a saturated plane map with no 3-ring.
}
\begin{lemma}
\label{lem:cc4e}
\lemcc
\end{lemma}

Using Lemma~\ref{lem:cc4e}, Thomassen \cite{rl:t86} rephrases Ungar's
Theorem A as follows.

\begin{lemma}
\label{lem:thomung}
Any cubic, cyclically 4-edge connected
planar graph
has a rectangular representation.
\end{lemma}

%Lemma \ref{lem:cc4e} justifies Thomassen's restatement
%of Ungar's Theorem A
%in Lemma \ref{lem:thomung}.

\subsection{New Characterization}
\label{ssec:new}

Ko\'{z}mi\'{n}ski and Kinnen \cite{rdual:kk85} define a {\em 4-triangulation} to be a
4-connected, planar triangulated graph with at least 6 vertices, at least
one of which has degree 4.
They prove \cite[Thm.~1]{rdual:kk85} that a
cube with one face that is a rectangular dissection is dual to a planar graph $G$
if and only if $G$ is a 4-triangulation.  
They also prove
\cite[Thm.~2]{rdual:kk85} that a planar graph $G$ with all faces
triangular except the outside has a rectangular dual if and only if $G$
can be obtained from some 4-triangulation $H$ by the deletion of some
degree-4 vertex and all its neighbors.  
We use the following somewhat weaker result to design
an algorithm for constructing rectangular layouts.

\begin{theorem}
\label{thm:gooddual}
If a planar graph $G$ can be derived
from some 4-connected planar triangulation
$H$ by the removal of some vertex and its incident edges,
then $G$ has a rectangular dual.
\end{theorem}
\begin{proof}
  Let $H$ be any 4-connected planar triangulation and $v$ any vertex
  of $H$.  By Lemma \ref{lem:ptc4ec}, $H^*$ is planar, cubic, and
  cyclically 4-edge connected, and so by Lemma \ref{lem:cc4e} $H^*$ yields
  a saturated plane map $M$ with no 3-ring.  Ungar's Theorem B
  \cite{rd:u53} implies that an isomorphic map $M'$ exists with the
  external face corresponding to $v$.  $M'$ is thus a rectangular dual for
  the graph derived from $H$ by removing $v$ and its incident edges.
\end{proof}

The benefit of
Theorem \ref{thm:gooddual} is that it gives a sufficient
condition for rectangular duality in terms of 4-connected triangulations
rather than
Ko\'{z}mi\'{n}ski and Kinnen's \cite{rdual:kk85} 4-triangulations.
This yields (in Section \ref{sec:graphs})
a simple augmentation procedure
for constructing rectangular layouts
based on the following alternative
characterization of graphs admitting rectangular layouts. 

\begin{theorem}
\label{thm:ft4ct}
A planar graph 
$G$ is a proper subgraph of a 4-connected planar triangulation
if and only if
$G$ has an embedding with no filled triangles.
\end{theorem}
\begin{proof}
($\Longrightarrow$) Let $G'$ be a 4-connected planar triangulation.  Let $G$ be the
  result of removing any one edge or vertex (and its incident edges)
  from $G'$.  By Lemma \ref{lem:4ctri},
  $G'$, and therefore $G$, has no separating triangle.  We claim that any embedding
  $\CE$ of $G$ has a non-triangular face or that $G$ is itself just a 3-vertex 
cycle.  To prove the claim, consider an arbitrary embedding $\CE'$ of $G'$. 
Removing a single edge from $\CE'$ yields an embedding $\CE$
(of $G$) with a non-triangular face.  If removing a vertex $v$ from 
$\CE'$ yields an embedding $\CE$ with no non-triangular face, then $\CE$ itself 
must be a simple triangle; otherwise, the triangular face of $\CE$ to which $v$
was adjacent is a 
separating triangle in $\CE'$, which by assumption cannot exist.  This proves
the claim.

If $G$ is a 3-vertex cycle, we are done; otherwise, by stereographic
  projection, we can assume that its external face is non-triangular.  Because any
  filled triangle is either a separating triangle or the external face,
  it follows that $\CE$ has no filled triangles, and therefore any proper subgraph
  of $G$ also has an embedding with no filled triangles.

($\Longleftarrow$)
  Now let $G$ be a planar graph, and let $\CE$ be some embedding of $G$ with
  no filled triangles.  Assume without loss of generality that $G$ has at
  least one non-triangular face.  Otherwise, $G$ itself is a
  triangulation, and the assumption that $\CE$ has no filled triangles
  implies that $G$ is simply a 3-vertex cycle.  We will show how to form
  by vertex augmentation a proper supergraph $G'$ of $G$ such that $G'$ is
  a planar triangulation with no separating triangles and hence by Lemma
  \ref{lem:4ctri} is 4-connected.
  
  First, we may assume that $G$ is biconnected.  Otherwise, we adapt
  a procedure attributed to Read \cite{biaug:r87}.  Consider
  any articulation vertex $v$, and let $u$ and $w$ be consecutive
  neighbors of $v$ in separate biconnected components.  Add new vertex
  $z$ and edges $\{z,u\}$ and $\{z,w\}$.  Iterating for every
  articulation point biconnects $G$ without adding separating triangles.
  Any face in the updated embedding $\CE$ is then bounded by a simple cycle,
  and the following procedure is well defined.

  Consider any non-triangular facial cycle $F$ in $\CE$.  Define a {\em
  chord} of $F$ to be a non-facial edge connecting two vertices of $F$.
  Consider any chord $\{x,y\}$ of $F$, and let $u$ and $v$ be the
  neighbors of $x$ on $F$.  There can be no edge $\{u,v\}$ in $G$, for
  such an edge would violate planarity.  Therefore embedding a new vertex
  $\nu(x)$ inside $F$ and adding edges $\{\nu(x),u\}$, $\{\nu(x),x\}$,
  and $\{\nu(x),v\}$ cannot create a separating triangle.  Let $F'$ be
  the new facial cycle defined by replacing the path $(u,x,v)$ in $F$
  by $(u,\nu(x),v)$, and iterate until $F'$ has no incident chords.
  Then, adding a final new vertex $\nu(F)$ with edges to each vertex
  on $F'$ completes the triangulation of the original face $F$ without
  creating separating triangles or modifying other faces.  Iterating for
  all non-triangular facial cycles completes the process, yielding a
  planar triangulation $G'$ with no additional separating triangles.

  Therefore, any separating triangle $T$ in $G'$ must have originally
  existed in $G$.  Because $\CE$ had no filled triangles, $T$ must
  have been embedded as a (triangular) face of $\CE$.  $T$ remains
  a face in $G'$, however, and because $G'$ is a triangulation,
  the removal of any face cannot disconnect $G'$, thereby contradicting
  the existence of $T$.
\end{proof}

\begin{corollary}
\label{cor:rlnft}
A graph $G$ has a rectangular layout
if and only if there exists some embedding of $G$
with no filled triangles.
\end{corollary}
\begin{proof}
Thomassen \cite[Thm.~2.1]{rl:t86} proves that $G$ has
a rectangular layout if and only if it is a proper subgraph
of a 4-connected planar triangulation.
The result then follows from Theorem \ref{thm:ft4ct}.
\end{proof}

\ignore{
The preceding vertex augmentation procedure
also
implies that if a graph $G$ can be derived from a 4-connected planar triangulation
$H$ by deleting one edge,
then $G$ can also be derived from a different 4-connected planar triangulation
$H'$ by deleting one vertex.
Hence,
by Theorem \ref{thm:gooddual},
a planar near-triangulation admits a rectangular dual
if and only if it is a vertex- or edge-maximal graph
that admits a rectangular layout.
}

Biedl, Kant, and Kaufman \cite{tri:bkk97} show how to transform
a planar embedding without separating triangles into a 4-connected
triangulation via edge augmentation, if possible.  
As they demonstrate, however,
it is not always possible to triangulate such a graph using
only edge augmentations.
Furthermore, we need the vertex-augmentation
method above for our algorithm in Section \ref{sec:graphs}.

\subsection{Rectangle of Influence Drawings}
\label{sec:const:rid}

Finally,
we link rectangular layouts to another graph visualization technique:
rectangle of influence drawings.
We use the definitions from Biedl, Bretscher, and Meijer \cite{rid:bbm99}
and Liotta et al.~\cite{rid:llmw98}.
A {\em (strong) closed rectangle of influence drawing} of a graph $G$
is a straight-line embedding of $G$ such that the
isothetic rectangular region, including the border, induced by any two vertices
$u$ and $v$
contains no other vertices if and only if $\{u,v\}$
is an edge in $G$.
A {\em weak closed rectangle of influence drawing}
relaxes the condition so that the
isothetic rectangular region, including the border,
induced by any two vertices
$u$ and $v$
contains no other vertices if $\{u,v\}$
is an edge in $G$.
A {\em (strong or weak) open rectangle of influence drawing}
is one in which all the isothetic rectangular interiors
obey the respective emptyness constraints;
the interiors of degenerate rectangles
are defined to be those of the induced line segments.
These drawings are also {\em planar}
if no two edges cross.

\begin{theorem}
A graph $G$ is a contact graph of rectangles
and thus admits a rectangular layout
if and only if $G$ has a planar, weak, closed rectangle of influence drawing.
\end{theorem}
\begin{proof}
This follows from 
Corollary \ref{cor:rlnft}
and 
Theorem 2 of Biedl, Bretscher, and Meijer \cite{rid:bbm99}.
\end{proof}

Rectangular layouts express the same class of
graphs under either weak or strong adjacency constraints,
but the same is not true of rectangle of influence drawings.
For example, a star on three (rsp., five) leaves has a 
planar, weak, open (rsp., closed) rectangle of influence drawing but 
no strong, open (rsp., closed) rectangle of influence drawing.
Liotta et al.~\cite{rid:llmw98}
characterize graphs with 
strong rectangle of influence drawings.
This settles an open problem
raised by Biedl, Bretscher, and Meijer \cite{rid:bbm99}.

\section{Layouts for General Graphs}
\label{sec:graphs}

Theorems \ref{thm:gooddual} and \ref{thm:ft4ct} suggest an algorithm for
constructing a rectangular layout for an arbitrary input graph $G$.
\begin{enumerate}
\item\label{st:emb} Construct an embedding $\CE$ of $G$ with no filled triangles.
        If no such embedding exists, then $G$ admits no
        rectangular layout.
\item\label{st:aug} Vertex-augment $\CE$ to create a proper supergraph $G'$ of $G$
        such that $G'$ is a 4-connected triangulation.
\item\label{st:dual} Construct a rectangular dual $R$ of $G''=G'-\{v\}$,
        where $v$ is any vertex added during the augmentation
        process in step \ref{st:aug}.
\item\label{st:gap} Replace each rectangle $r$ in $R$ that corresponds to a vertex
        added during step \ref{st:aug} by a gap.
        The result is a rectangular layout for $G$.
\end{enumerate}

\begin{theorem}
\label{thm:grapharea}
An $O(n^2)$-area rectangular layout can be built
in $O(n)$ time
for any contact graph $G$ of rectangles.
If $G$ is not a contact graph of rectangles,
this can be discovered in $O(n)$ time.
\end{theorem}
\begin{proof}
  Biedl, Kant, and Kaufmann \cite[Thm.~5.5]{tri:bkk97} show how to
  construct an embedding $\CE$ of $G$ with no filled triangles if one
  exists, or detect if no such embedding exists, both in $O(n)$ time.
  
  The proof of Theorem \ref{thm:ft4ct} outlines a procedure to effect
  step \ref{st:aug}.  Finding articulation points can be done in $O(n)$
  time by depth-first search \cite{algs:ahu}.  Representing $\CE$ by a
  standard doubly connected edge list \cite{dcel:mp78} then allows all
  operations to be implemented in $O(n)$ time overall.  In particular,
  after augmenting to assure biconnectivity, iterating over the faces of
  $\CE$ takes $O(n)$ time plus the time to process each face.  Iterating
  over the vertices of all the faces takes $O(n)$ time plus the time
  to process each vertex.  Processing each vertex $x$ on each face $F$
  involves checking each incident edge $e$ to see if $e$ is a chord of
  $F$; each such test takes $O(1)$ time, and each edge in $G$ is checked
  twice, once for each endpoint, for a total of $O(n)$ time.  If $e$
  is a chord, augmenting $F$ to replace $x$ with $\nu(x)$ also takes
  $O(1)$ time.  Adding vertex $\nu(F)$ takes time linear in the number
  of vertices on $F$; over all faces this is $O(n)$ time.  In all, step
  \ref{st:aug} can be done in $O(n)$ time, yielding graph $G'' \supseteq
  G$ with $O(n)$ vertices.
  
  Theorem \ref{thm:gooddual} asserts that $G''$ has a rectangular dual.
  He \cite{rdual:h93} shows how to construct an $O(n^2)$-area rectangular
  dual of $G''$ in $O(n)$ time.\footnote{
	He does not explicitly state the area 
	of the dual resulting from his construction,
	but the bound is easily derived.}
  During the construction, we simply
  indicate that any rectangle corresponding to a vertex in $G''-G$ should
  instead be rendered as a gap.  Since each edge in $G''-G$ is incident to
  at least one vertex in $G''-G$, the result is a rectangular layout for
  $G$.
\end{proof}

\subsection{General Lower Bound}

A trivial, worst-case lower bound for graphs is 
$
\max\left\{
        n,
        \sum_{v\in V} \left\lceil\frac{\dg(v)}{4}\right\rceil,
        \sum_{v\in V} \left\lceil\frac{\dg(v)-2}{2}\right\rceil
\right \},
$
where $\dg(v)$ is the degree of vertex $v$.
The second term comes from the fact that each vertex
$v$ is represented by a rectangle, which has 4 sides;
the area of that rectangle must therefore be at least
$\lceil\frac{\dg(v)}{4}\rceil$ to accommodate all the
adjacencies.
This is tight in general:  consider the infinite grid,
in which each vertex has degree 4 and the area required is $|V|$.
The third term generalizes this argument.
The perimeter of $v$'s rectangle must be at least $\dg(v)$ units.
If the sides of the rectangle have lengths $a$ and $b$,
then minimizing $ab$ subject to $a+b\geq d/2$
yields that $ab\geq \lceil\frac{\dg(v)-2}{2}\rceil$.

To show a worst-case lower bound that matches our upper bound, first 
define an {\em $n$-rung ladder} to be a graph on at least
$n+2$ vertices---$L$, $R$, and $x_i$ for $1\leq i\leq n$---with
edges
$\{L,x_i\}$ and
$\{x_i,R\}$ for $1\leq i \leq n$
and paths (possibly including additional vertices)
connecting $x_i$ to $x_{i+1}$ for $1\leq i < n$.
We call the $x_i$'s the {\em rungs} and $L$ and $R$ the {\em struts} of the ladder.

In a rectangular layout,
call some rectangle $r$ {\em above} some rectangle $s$
if the lowest extent of $r$ is 
no lower than the highest extent of $s$.
Symmetrically define {\em below}, {\em right of}, and {\em left of}.
Call a set of rectangles {\em vertically} (rsp., {\em horizontally}) {\em stacked}
if their above (rsp., left of) relationships form a total order.
A set of rectangles is {\em vertically} (rsp., {\em horizontally}) {\em aligned}
if they have pairwise identical projections onto the $x$-axis (rsp., $y$-axis).
We use the {\em length} of a rectangle to mean the maximum
of its width and height.

\begin{figure}[t]%
	\begin{center}
	\input{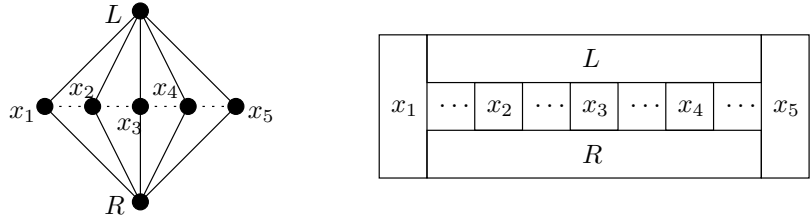}
	\end{center}
	\caption{A 5-rung ladder (dotted lines indicate
	paths) and a possible layout.
	In any layout,
	all the $x_i$ must be stacked, and
	the rectangles $x_2$, $x_3$, and $x_4$ must be 
	aligned as shown between $L$ and $R$.
	The only other variations also align rectangle(s)
	$x_1$ and/or $x_5$, but the width remains at least 5.
	}%
	\label{fig:ladder}%
	\end{figure}

\begin{lemma}
\label{lem:ladder}
Assume $n\geq 3$.
Any rectangular layout for an $n$-rung ladder
must possess
one of the following sets of properties:
\begin{enumerate}
\item
width at least $n$ and height at least 3;
rectangles $x_i$ for $1\leq i\leq n$
all horizontally stacked;
and rectangles $x_i$ for $1< i < n$
all horizontally aligned
between $L$ and $R$; or
\item
height at least $n$ and width at least 3;
rectangles $x_i$ for $1\leq i\leq n$
all vertically stacked;
and rectangles $x_i$ for $1< i < n$
all vertically aligned
between $L$ and $R$.
\end{enumerate}
\end{lemma}
\begin{proof}
Consider the path $(x_1,S_1,x_2,S_2,\ldots,x_{n-1},S_{n-1},x_n)$,
where the $S_i$'s are possibly null paths,
connecting the $x_i$'s.
We prove the lemma by induction on the total number of
vertices in the $S_i$'s.
Refer to Figure \ref{fig:ladder}.
We interchange the notion of vertices and rectangles
and rely on context to disambiguate.

The base case is when all the $S_i$'s are null; i.e.,
there is a direct path $(x_1,x_2,\ldots,x_n)$.
Let $L$ be placed above $R$; other cases are symmetric.
Only $x_1$ and $x_n$ may be to the sides (left and right)
of $L$ and $R$,
for
if a different $x_i$ were, say, to the left of $L$ and $R$,
abutting both,
then one of $x_{i-1}$ and $x_{i+1}$ would not be able
to abut both $L$ and $R$.
Thus, $x_2,\ldots,x_{n-1}$ must be below $L$ and above $R$;
that they must each abut both $L$ and $R$
therefore implies that
these rectangles must be horizontally aligned.
If $x_1$ and/or $x_n$ are to the sides of $L$ and $R$,
all the $x_i$'s are stacked; if $x_1$ and $x_n$ are also
below $L$ and above $R$, all the rectangles are horizontally aligned
and hence also horizontally stacked.
That the rectangles are horizontally stacked implies that the width is $n$.
The height follows by construction.
This proves the base case.

Given a layout for any graph,
removing the rectangle corresponding to some vertex---i.e.,
turning it into a gap---must
produce a layout for the corresponding proper subgraph.
This proves the inductive step.
\end{proof}

Define an {\em $(i,j)$-ladder} to be a graph on $i+j+2$
vertices:
an $i$-rung ({\em external}) ladder defined by some vertices
$L$, $R$, and $x_k$ for $1\leq k\leq i$,
united with a $j$-rung ({\em internal}) ladder
defined by $x_{\lfloor i/2 \rfloor}$, $x_{\lfloor i/2 \rfloor +1}$,
and $y_k$ for  $1\leq k\leq j$.
See Figure \ref{fig:graphlb}.
\fig{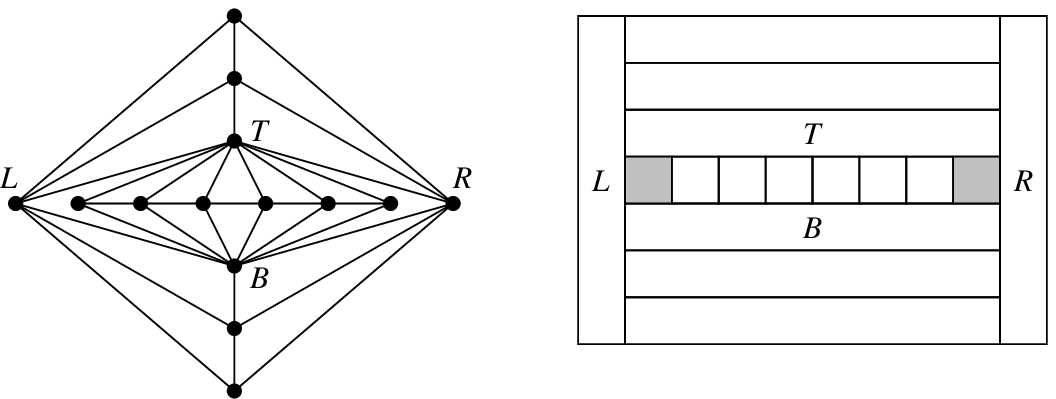}{A $(6,6)$-ladder and a possible layout.}{fig:graphlb}

\begin{theorem}
For $n\geq 4$, any layout for
an $(n,n)$-ladder has area $\Omega(n^2)$.
\end{theorem}
\begin{proof}
Lemma \ref{lem:ladder} applied to the 
external ladder shows that the height (or, rsp., width)
of any layout is at least $n$
and that the rectangles 
$x_{\lfloor n/2 \rfloor}$ and $x_{\lfloor n/2 \rfloor +1}$
are vertically (or, rsp., horizontally) aligned between $L$ and $R$.
Say the height is $n$; the other case is symmetric.
Then Lemma \ref{lem:ladder} shows that the width
induced by the internal ladder is at least $n$.
The theorem follows.
\end{proof}

\section{Layouts for Trees}
\label{sec:trees}

We present an algorithm that constructs $O(n\log n)$-area 
rectangular layouts for trees.
We then show a matching worst-case lower bound.
There do exist trees with better layouts, however,
and we constructively show an infinite class of trees
that have $O(n)$-area layouts.
As with general graphs,
this leaves open the problem of devising better approximation
algorithms for trees.

\subsection{General Algorithm}

Given an undirected tree, $T=(V,E)$,
assume $T$ is rooted at some vertex $r$; if not, pick an arbitrary root.
A simple Algorithm A lays out $T$ as follows.
For all $v\in V$,
let $\desc(v)$ be the number of descendants of $v$.
Each vertex $v$ is represented as a rectangle of height 1
and width $\desc(v)$.
For any $v$, the rectangles for its children are placed
under the rectangle for $v$.
The rectangle for the root appears on top.
For simplicity, we allow corners to meet trivially;
with our construction, we can eliminate this problem at a constant-factor
area penalty.
Similarly, we ignore the issue of strong versus weak layouts.
See Figure \ref{fig:trees}(a).

\fig{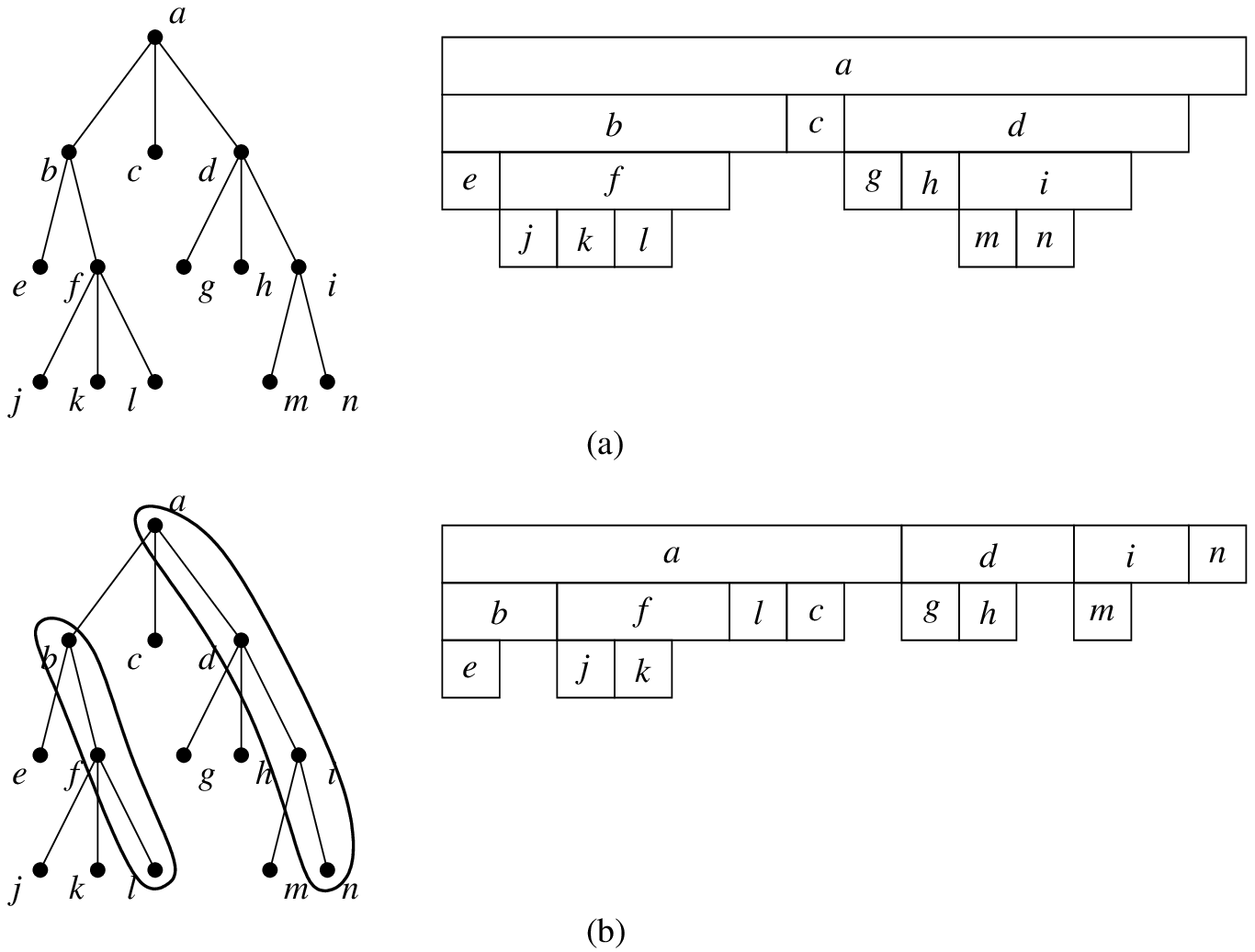}{(a) A tree and the result of applying Algorithm A.
The width of each rectangle is the number of descendants of the
corresponding vertex; e.g., the width of $f$ is 4, and the width of $d$ is 6.
(b) A partition of the tree into paths---uncircled 
nodes form singleton paths---and the application of Algorithm B to
the partition.
}{fig:trees}

\begin{lemma}
Algorithm A produces a layout of $T$ of area $n\cdot\depth(T)$.
\end{lemma}
\begin{proof}
The assertion about area is straightforward.
Correctness follows by induction from the fact that each rectangle
is wide enough to touch all the rectangles
for its vertex's children plus one unit for itself.
\end{proof}

Consider a partition of $T$ into a collection of vertex-disjoint paths.
Algorithm A generalizes
into Algorithm B by abutting all rectangles in a single path
horizontally and abutting rectangles for deeper children in the partition vertically.
Details follow.
Refer to Figure \ref{fig:trees}(b).

For a path $P=(u_0,\ldots,u_k)$,
in top-down order, define $t(P) = u_0$,
i.e., the vertex in $P$ of maximum height.
Define $\desc(P) = \desc(t(P))$.
Define $s(u_k) = \desc(u_k)$,
and for $0\leq i < k$,
define $s(u_i) = \desc(u_i) - \desc(u_{i+1})$.
Note that $\sum_{x\in P} s(x) = \desc(P)$.

Each path $P=(u_0,\ldots,u_k)$ is represented as 
a rectangle of height 1 and width $\desc(P)$,
which is partitioned into rectangles of
width
$s(u_0),\ldots,s(u_k)$,
each representing the corresponding $u_i \in P$.
The rectangle for path $P$ with $t(P) = r$
is placed on top.
Under the rectangle for each $u_i\in P$
(for each $P$)
are placed the rectangles for each path $P'$
such that $t(P')$ is a child of $u_i$.

\begin{lemma}
\label{lem:layout}
Algorithm B produces a layout of $T$.
\end{lemma}
\begin{proof}
Consider any vertex $x$.
We need only account for the adjacency $(x,p(x))$
(assuming $x\not=r$),
because any child $u$ of $x$ is accounted for
by the adjacency $(u,p(u))$.

Denote by $B(x)$ the rectangle representing $x$.
If $x\not=r$, then $B(x)$ overlaps $B(p(x))$,
for either $x$ and $p(x)$ are in a common path,
in which case their rectangles share a vertical side,
or else $B(x)$ is layed out underneath $B(p(x))$.
The construction assures that $B(p(x))$ is wide enough
in this latter case.
\end{proof}

Consider the {\em compressed tree} $C(T)$,
formed by compressing each path $P$
into a super-vertex,
with edges to each super-vertex $P'$
such that $t(P')$
is a child of some $x\in P$.

\begin{lemma}
\label{lem:depth}
Algorithm B produces a layout of $T$
of area $n\cdot\depth(C(T))$.
\end{lemma}
\begin{proof}
Algorithm B produces a one-unit high collection
of rectangles
for each distinct depth in $C(T)$.
Each such collection is of width no more than $n$
(the total number of descendants of the paths
at that depth).
\end{proof}

We use the heavy-path partition of $T$,
as defined by Harel and Tarjan \cite{nca:ht}
and later used by Gabow \cite{nca:g}.\footnote{
Tarjan \cite{pathcomp:t} originally introduced heavy-path partitions,
but defined in different terms;
Schieber and Vishkin \cite{lca:sv} later used yet another variant.}
Call tree edge $(v,p(v))$ {\em light} if $2 \cdot\desc(v)  \leq \desc(p(v))$,
and {\em heavy} otherwise.  
Since a heavy edge must carry more than half the descendants of a vertex, each
vertex can have at most one heavy edge to a child,
and therefore deletion of the
light edges produces
a collection of vertex-disjoint {\em heavy paths}.
(A vertex with no incident heavy edges
becomes a singleton, called a {\em trivial heavy path}.)

\begin{theorem}
Algorithm B applied to the heavy-path partition of $T$
produces a layout of area $O(n\log n)$ in $O(n)$ time.
\end{theorem}
\begin{proof}
The compressed tree, $C(T)$, is constructed
by contracting each heavy
path in $T$ into a single super-vertex.
Each tree edge in $C(T)$ corresponds to a light edge of $T$.
Since there are $O(\log n)$ light edges on the path from any vertex to the root of $T$,
$C(T)$ has depth $O(\log n)$.
The area bound follows from Lemmas \ref{lem:layout} and \ref{lem:depth}.
$C(T)$ can be built in $O(n)$ time after a depth-first search;
the rest of the algorithm performs $O(1)$ work per vertex.
\end{proof}

\subsection{General Lower Bound}

We show there exists an infinite family of trees that require
$\Omega(n\log n)$ area for any layout.  First we show that
any layout of a binary tree has $\Omega(\log n)$ length
in each dimension.

We define the notion of paths in layouts.
A {\em path} in layout $L$ is a sequence $(r_1,\ldots,r_{\ell})$
of rectangles in $L$ such that for each $1\leq i<\ell$,
the boundaries of $r_i$ and $r_{i+1}$ overlap.
A path in a strong layout thus corresponds to a path
in the underlying graph.
A {\em vertical extremal path}
of $L$ is a path that
touches both the top and bottom of $L$'s bounding box $B$;
similarly, define a {\em horizontal extremal path}
to touch the left and right sides of $B$.
An extremal path that touches opposite corners of $B$
is both vertical and horizontal.
By definition, every layout has at least one vertical and at least one horizontal
extremal path, possibly identical.

Consider graph $G$ with some layout $L$
and some subgraph $G'$.
$L$ contains a sub-layout $L'$ for $G'$.
Any extremal path in $L'$ induces a path in $L$.
Two sub-layouts are {\em disjoint}
if their induced subgraphs are disjoint.
Extremal paths for disjoint sub-layouts may not cross in $L$,
for this would imply two non-disjoint rectangles,
a fact codified as follows.

\begin{fact}
\label{fac:nocross}
Consider graph $G$, some layout $L$ of $G$,
and any two disjoint, connected subgraphs $G_1$ and $G_2$ of $G$.
Extremal paths for the corresponding sub-layouts $L_1$ and $L_2$ may not cross in $L$.
\end{fact}

\begin{lemma}
\label{lem:overlay}
Let $G_1$ and $G_2$ have minimal area layouts
with length at least $d$ in each dimension.
Let $G$ contain both $G_1$ and $G_2$ as subgraphs.
Then any layout $L$ of $G$ has length at least $d+1$
in at least one dimension.
\end{lemma}
\begin{proof}
Since $L$ contains sub-layouts $L_1$ and $L_2$ for $G_1$ and $G_2$,
rsp.,
by assumption
$L$ has length at least $d$ in both dimensions.
If $L$ has width and height both $d$,
then any horizontal extremal path for $L_1$
must cross any vertical extremal path for $L_2$,
contradicting Fact \ref{fac:nocross}.
Thus, $L$ must have width or height at least $d+1$.
(See Figure \ref{fig:nocross}.)
\end{proof}
\fig{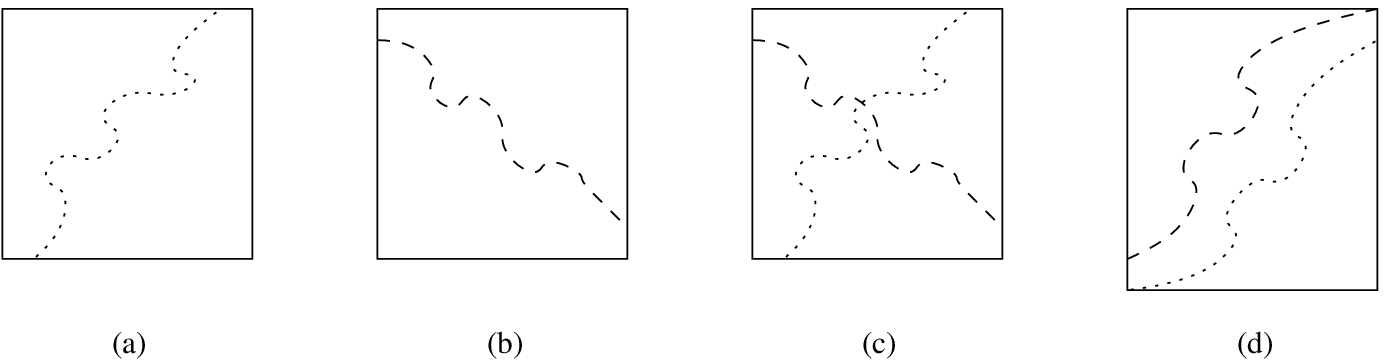}{Bounding boxes and extremal paths for various
        layouts.  (a) Layout $L_1$ for $G_1$;
        (b) Layout $L_2$ for $G_2$;
        each of size $d$-by-$d$.
        (c) Impossible $d$-by-$d$ layout for $G$ containing $G_1$ and $G_2$
                as subgraphs; any corresponding extremal paths
                would have to cross.  
        (d) $d$-by-$(d+1)$ layout for $G$.  The extremal path for $L_1$
                is dotted, and that for $L_2$ is dashed.
                By extending $G$ in one dimension, the extremal paths
                need not cross.}{fig:nocross}

\begin{figure}[t]%
	\begin{center}
	\input{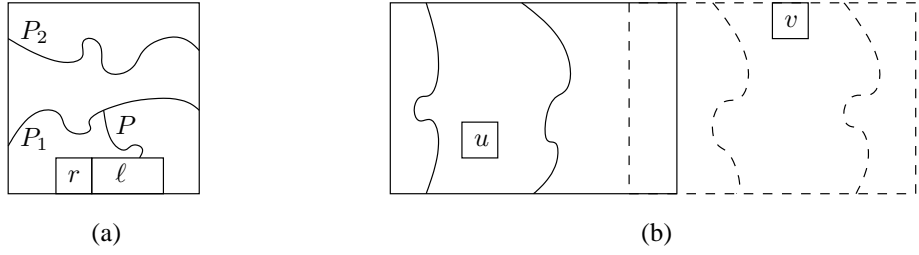}
	\end{center}
	\caption{(a) A bounding box of width $d$ containing a layout $L$
        that includes two sub-layouts $L_1$ and $L_2$,
        each of minimum dimension $d$.
        If $r$ is adjacent to the bottom of $L$, let $\ell$ be a rectangle
        in $L_1$ adjacent to $r$.
        There is some path $P$ in $L_1$ connecting $\ell$ to a horizontal
        extremal path $P_1$ of $L_1$ and $L$.
        No horizontal extremal path of $L_2$ can intersect the closed
        curves formed by $P$, $P_1$, and the bounding box of $L$.
        Thus $P_2$ is above $P_1$, but then $r$ cannot be connected
        to $P_2$ without crossing $P_1$.
        (b) Bounding boxes for layouts of two $T_{i-1}$'s, 
        rooted at $u$ (solid) and $v$ (dashed).
        If both layouts are of height $\left\lfloor \frac{i-2}{2} \right\rfloor$,
        then extremal paths (dashed for the $T_{i-1}$ rooted at $v$) 
        separate $u$ from $v$,
        and $r$ cannot be placed within this height.}%
	\label{fig:jordan}%
	\end{figure}

\begin{lemma}
\label{lem:middle}
Let $G_1$ and $G_2$ be graphs such that all their layouts have 
length at least $d$ in each dimension.
Let $G$ be formed by adding a new vertex $r$,
adjacent to one vertex in each of $G_1$ and $G_2$.
Then in any layout $L$ of $G$ with length $d$ in some dimension,
$r$ cannot be incident to a length-$d$ side of the bounding box
of $L$.
Furthermore there exist extremal paths in the length-$d$ dimension
to either side of $r$.
\end{lemma}
\begin{proof}
(Refer to Figure \ref{fig:jordan}(a).)
Assume to the contrary that $L$ has width $d$
and $r$ is adjacent to the bottom (sym., top) of $L$;
the argument for height $d$ is symmetric.
$L$ induces layouts $L_1$ of $G_1$ and $L_2$ of $G_2$,
each by assumption with length at least $d$ in each dimension $d$.
Let $r$ be adjacent to a rectangle $\ell$ of $L_1$ (sym., $L_2$).
Then there must be a path $P$ from $\ell$ that intersects
a horizontal extremal path $P_1$ of both $L_1$ and $L$.
The union of $P$ and one side of this extremal path
forms a closed curve with the bounding box that contains $r$.

Now consider a horizontal extremal path $P_2$ of $L_2$,
which is also a horizontal extremal path of $L$.
There must be a path $P'$ in $L_2$ connecting $P_2$
to $r$.
$P_2$ cannot intersect the closed curve defined above
without creating non-disjoint rectangles;
thus $P_2$ is above $P_1$.
But then by the Jordan curve theorem \cite[Section 8-13]{top:m99},
$P'$ itself must cross the curve to reach $P_2$,
which again would create non-disjoint rectangles.

If $r$ is above $P_2$, a similar contradiction holds.
Thus, $r$ must be between $P_1$ and $P_2$.
\end{proof}

Define $T_i$ to be a complete binary tree on $2^i$ leaves.

\begin{lemma}
\label{lem:treedim}
Any layout for $T_i$ has length at least $\lfloor i/2 \rfloor$ in each dimension.
\end{lemma}
\begin{proof}
The theorem is true for $i=0$ and $i=1$.  
Assuming it is true up to $i-1$, we prove
it by induction for $i\geq 2$.
Denote by $r$ the root of $T_i$ and by $u$ and $v$ the roots
of the $T_{i-1}$'s rooted at the children of $r$.
By induction, each $T_{i-2}$ rooted at children of $u$ and $v$
has length at least $\left\lfloor \frac{i-2}{2} \right\rfloor$ in each dimension.
By Lemma \ref{lem:overlay} therefore, the layouts for the $T_{i-1}$
subtrees rooted at $u$ and $v$
each have at least one dimension of length 
$\left\lfloor \frac{i-2}{2} \right\rfloor + 1 = \lfloor i/2 \rfloor$.
If either sub-layout has both dimensions this large, we are done.
If one sub-layout has width $\lfloor i/2 \rfloor$ and the other height $\lfloor i/2 \rfloor$,
we are similarly done.

Therefore, assume the sub-layouts for both $T_{i-1}$'s
have height $\left\lfloor \frac{i-2}{2} \right\rfloor$.
(Symmetrically argue if both widths are this small.)
Also assume that in the layout for $T_i$,
$r$ is not placed on top of the two sub-layouts,
or the height grows by the required one unit.
By Lemma \ref{lem:middle},
neither $u$ nor $v$ may be adjacent to the left or right side
of their layouts,
and furthermore, there are vertical extremal paths
due to their own children that separate $u$ from $v$.
But then by Jordan curve arguments,
there cannot be paths connecting both $u$ and $v$
to $r$ without violating rectangle disjointedness.
(See Figure \ref{fig:jordan}(b).)
\end{proof}

Assume $n=2^k$ for some integer
$k\geq 1$.
Let $S_n$ be a tree formed by linking the root of $T_k$
to the root of a star with $n$ leaves.

\begin{theorem}
\label{thm:treelb}
Any layout $L_n$ for $S_n$ has $\Omega(n\log n)$ area.
\end{theorem}
\begin{proof}
$L_n$ includes sub-layouts for $T_k$ and the star on $n$ leaves.
Lemma \ref{lem:treedim} implies that the length
of each dimension
of $L_n$ is $\Omega(\log n)$.
The only layout for a star with $n$ leaves is a rectangle for the root
with $n$ rectangles around its perimeter.  
Thus, at least one dimension of the star layout is
of length $\Omega(n)$.
The theorem follows.
\end{proof}

\subsection{Linear Area for Complete Trees}

Let $T_n^k$ be a complete tree of arity $k$ on $n$ leaves
and $L_n^k$ a corresponding strong layout.
$L_1^k$ is the unit square.
For higher $n$,
construct $L_n^k$ as follows.
Recursively construct the $k$ $L_{n/k}^k$ sub-layouts,
assuming the root of each one has unit length in one dimension.
Attach the roots of the sub-layouts by their unit-length sides
to a root rectangle of height 1 and appropriate width,
leaving
one unit of horizontal space to the right of each sub-layout.
See Figure \ref{fig:opttree}.
\fig{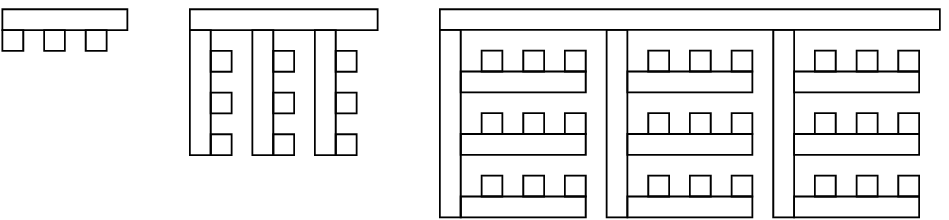}{Layouts $L^3_3$ (left), $L^3_9$ (middle),
        and $L^3_{27}$ (right).}{fig:opttree}

\begin{theorem}
\label{thm:opttree}
$L_n^k$ is of area $O(n)$.
\end{theorem}
\begin{proof}
For clarity, we drop the superscripts.
Denote by $h_n$ the height and $w_n$ the width of $L_n$.
Then $h_1 = w_1 = 1$;
$h_k = 2$, $w_k = 2k$;
and for higher $n$, we have the following \emph{type-1 recurrence}:
\begin{eqnarray*}
h_n & = &  1+ w_{n/k}; \\
w_n & = & kh_{n/k} + k.
\end{eqnarray*}
Simplifying:
\begin{eqnarray*}
h_n & = & kh_{n/k^2} + k+1; \\
w_n & = & kw_{n/k^2} + 2k.
\end{eqnarray*}
Solving these recurrences for $n=k^c$ yields 
$$
h_n = \Theta\left(k^{\lfloor c/2 \rfloor}\right);
w_n = \Theta\left(k^{\lceil c/2 \rceil}\right).
$$
Therefore, the area, which is $h_n w_n$, is $\Theta(n)$.
\end{proof}

An alternative layout yields a family of layouts of linear area,
but with elastic widths and heights.
Rather than
attaching the roots of the sub-layouts to the root of the layout by
their unit-length sides, 
consider attaching them by their other sides. 
This establishes a
recurrence of the form
\begin{eqnarray*}
h_n & = &  1+ h_{n/k}; \\
w_n & = & kw_{n/k} +k;
\end{eqnarray*}
which yields the solution $h_n = \Theta(\log_k n), w_n = \Theta(n)$. We
call this a \emph{type-2 recurrence}.

Combining the above two recurrences, we can prove the following.

\begin{theorem}
\label{thm:reconfig}
For any constant $\alpha\in [\frac{1}{2}, 1)$, a layout
for $T_n^k$ exists with width $\Theta(n^\alpha)$ and height $\Theta(n^{1-\alpha})$. 
\end{theorem}
\begin{proof}
If we apply a type-1 recurrence $2m$ times, we obtain the recurrence
\begin{eqnarray*}
h_n & = & (1+k)(1 + k + \cdots + k^{m-1}) + k^mh_{n/k^{2m}}; \\
w_n & = & 2k(1+k+\cdots + k^{m-1}) + k^m w_{n/k^{2m}};
\end{eqnarray*}
which simplified yields
\begin{eqnarray*}
h_n & = & \Theta(k^m) + k^m h_{n/k^{2m}}; \\
w_n & = & \Theta(k^m) + k^m w_{n/k^{2m}}.
\end{eqnarray*}

Similarly, a type-2 recurrence applied $2\ell$ times yields 
\begin{eqnarray*}
h_n & = &  2\ell+ h_{n/k^{2\ell}}; \\
w_n & = & k^{2\ell} + k^{2\ell} w_{n/k^{2\ell}}.
\end{eqnarray*}

Consider a layout in which we apply a type-1 construction $2m$
times and then a type-2 construction $2\ell$ times and then repeat. The
recurrence governing this construction is given by a combination of the
above two recurrences:
\begin{eqnarray*}
h_n & = & \Theta(k^m) + k^m (2\ell+ h_{n/k^{(2\ell+2m)}}); \\
w_n & = & \Theta(k^m) + k^m (k^{2\ell} + k^{2\ell} w_{n/k^{(2\ell+2m)}});
\end{eqnarray*}
which simplifies to
\begin{eqnarray*}
h_n & = & \Theta(k^m \ell) + k^m h_{n/k^{(2\ell+2m)}}; \\
w_n & = & \Theta(k^{m+2\ell}) + k^{m+2\ell} w_{n/k^{(2\ell+2m)}}.
\end{eqnarray*}

Solving these recurrences and setting $\alpha = 1 - m/(2m+2\ell)$, we get
$h_n = \Theta(n^{1-\alpha}(1+\ell))$ and $w_n = \Theta(n^\alpha)$. For any $1/2
\le \alpha < 1$ we can find constants $\ell$ and $m$ 
that satisfy this equation. 
\end{proof}

\section{Area Optimality}
\label{sec:area}

\subsection{NP-Hardness of Generating Optimal Layouts}

Recall the problem of {\em numerical matching with target sums}
\cite{gandj:gj}.
Given are disjoint sets $X$ and $Y$, each of $m$ elements,
a size $s(a)\in{\mathbb Z}^+$ for each $a \in X \cup Y$,
and a target vector $B=(B_1,\ldots,B_m)$
with each $B_i\in{\mathbb Z}^+$.
The problem is to determine if $X\cup Y$
can be partitioned into $m$ disjoint sets $A_1,\ldots,A_m$,
each containing exactly one element from each of $X$ and $Y$,
such that for $1\leq i\leq m$, $\sum_{a\in A_i} s(a) = B_i$.
The problem is strongly NP-hard in general.
Consider some instance \II~of numerical matching with target sums.
Assume without loss of generality
that 
\begin{equation}
\label{eq:iiass}
\sum_{x\in X} s(x) + \sum_{y\in Y} s(y) = \sum_{i=1}^m B_i,
\end{equation}
or else \II~has no solution.
We will construct a graph $G(\II)$ that has an optimal layout of certain dimensions
if and only if $\II$ has a solution.

Define an {\em $n$-accordion} to be a graph on $3n+2$
vertices:
three disjoint, simple paths of length $n$ each;
two additional vertices $x$ and $y$;
an edge between $x$ and each vertex on the first and second paths;
and an edge between $y$ and each vertex on the second and third paths.
An {\em enclosed accordion} is an accordion augmented ({\em enclosed})
by two additional
vertices $T$ and $B$:
$T$ adjacent to each vertex on the first path,
and $B$ adjacent to each vertex on the third path.
See Figure \ref{fig:accord}(a).
\fig{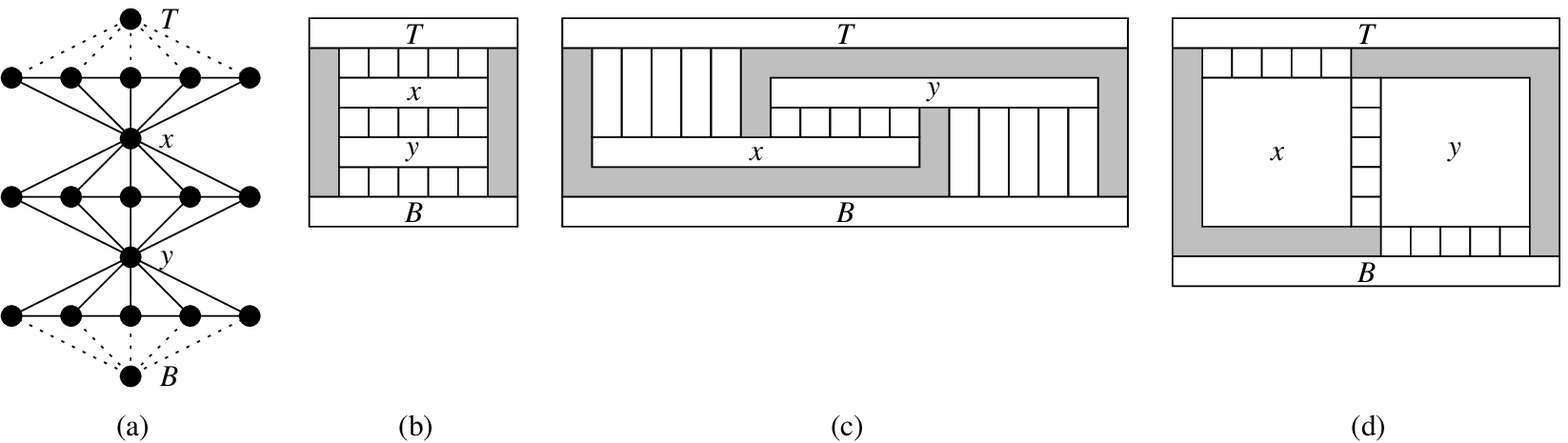}{(a) A 5-accordion enclosed by $T$ and $B$.
	The accordion edges are solid; the enclosing edges are dotted.
	(b)--(d) Layouts with
		(b) $y$ below $x$;
		(c) $x$ below $y$;
	and	(d) neither $x$ nor $y$ below the other.
}{fig:accord}

\begin{lemma}
\label{lem:enclosedaccord}
Assume $n\geq 3$.
In any layout of an enclosed $n$-accordion
such that $T$ appears above (sym., below) $B$
and no rectangle corresponding to an accordion vertex
is left or right of $T$ or $B$,
the accordion rectangles form a bounding box of height at least $5$
and width at least $n$;
furthermore, to achieve height $5$ and width $n$ simultaneously,
the accordion rectangles must form a dissection.
Symmetrically, 
in any layout of an enclosed $n$-accordion
such that $T$ appears left (sym., right) of $B$
and no rectangle corresponding to an accordion vertex
is above or below $T$ or $B$,
the accordion rectangles form a bounding box of height at least $n$
and width at least $5$;
furthermore, to achieve height $n$ and width $5$ simultaneously,
the accordion rectangles must form a dissection.
\end{lemma}
\begin{proof}
We prove the case in which $T$ is above $B$; the other cases are symmetric.
Refer to Figure \ref{fig:accord}.
The width lower bound follows, because all vertices
on the first path must be adjacent to $T$'s bottom boundary.

By assumption,
in any layout $x$ must be below $T$ and $y$ above $B$.
Lemma \ref{lem:ladder} implies that the width or height of the ladder
between $x$ and $y$
must be at least $n$.
If the height is at least $n$ (as in Figure \ref{fig:accord}(d)),
the height lower bound follows
from the mutual non-adjacency of $x$, $y$, $T$, and $B$
and the lower bound on $n$.
If the height of the $x$-$y$ ladder is less than $n$,
then Lemma \ref{lem:ladder} implies that the width must be
at least $n$ and the height at least 3.
Because neither $x$ nor $y$ can abut $T$ or $B$,
there must be at least one additional unit of height each above and below
the $x$-$y$ ladder to connect it $T$ and $B$ via the intermediate vertices.
Example configurations are depicted in Figures \ref{fig:accord}(b)--(d).
Thus the overall height of the accordion must be at least 5.

To achieve height $5$ and width $n$ simultaneously,
the $x$-$y$ ladder itself must be of height 3, 
by the same argument that 2 additional units of height
are required to connect it to $T$ and $B$.
By Lemma \ref{lem:ladder}, therefore, the $x$-$y$ ladder must have width at least $n$,
with the rectangles other than $x$ and $y$ stacked and the middle ones aligned
horizontally.
If they were simply stacked, however, then the width of $x$ and $y$
would be only 3, which would not suffice to place the rectangles
between them and $T$ and $B$.  Hence {\em all} of the rectangles of the $x$-$y$
ladder other than $x$ and $y$ must be aligned, which then implies that the rectangles between
$x$ and $T$ and those between $y$ and $B$ must also be aligned,
as shown in Figure \ref{fig:accord}(b), to meet the overall width assumption.
This forms a dissection, as claimed.
\end{proof}

\begin{figure}[t]%
	\begin{center}
	\input{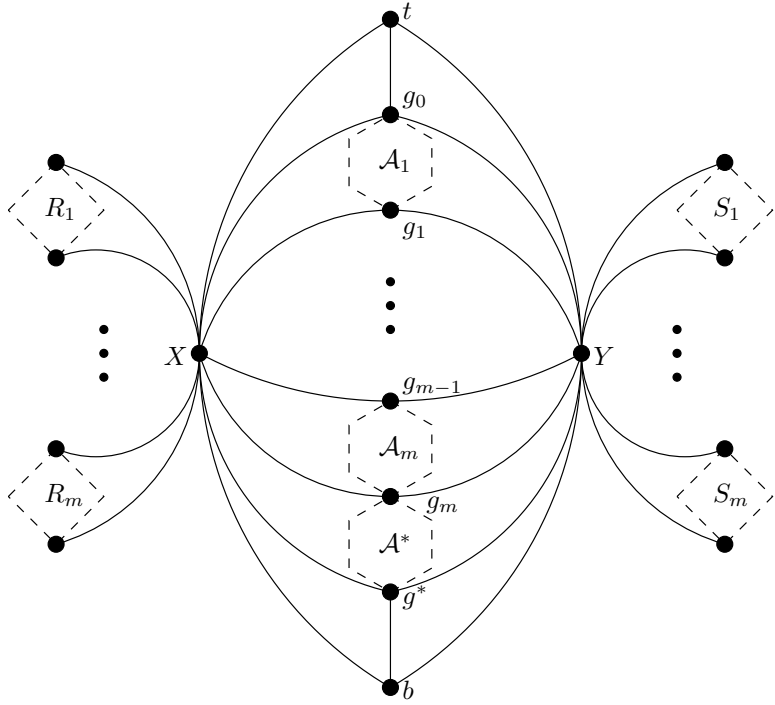}
	\end{center}
	\caption{The graph $G(\II)$ for some instance $\II$ of
	numerical matching with target sums.  The accordions
	are denoted by dashed hexagons and the ladders by dashed diamonds.
}%
	\label{fig:npconst}%
	\end{figure}

Let $x_1,\ldots,x_m$ (rsp., $y_1,\ldots,y_m$)
denote the sizes of the elements of $X$ (rsp., $Y$) in \II;
define $\bmax=\max\{ B_i : 1\leq i\leq m\}$;
and define $\Delta_i = 2\bmax - B_i \geq 1$ for $1\leq i\leq m$.
Graph $G(\II)$ is formed
from the following components.
\begin{itemize}
\item vertices $X$, $Y$, $t$, $b$, $g^*$, and $g_i$ for $0\leq i\leq m$;
\item an $x_i$-rung ladder $R_i$
	and a $y_i$-rung ladder $S_i$ for $1\leq i\leq m$;
\item a $\Delta_i$-accordion, denoted $\CA_i$, for $1\leq i\leq m$;
\item a $(2\bmax+2)$-accordion, denoted $\CA^*$.
\end{itemize}
The components are arranged as follows.
(See Figure \ref{fig:npconst}.)
\begin{itemize}
\item $X$ and $Y$ are each adjacent to 
	$t$, $b$, $g^*$ and $g_i$ for $0\leq i\leq m$;
\item $t$ is adjacent to $g_0$ and $b$ to $g^*$;
\item $g_{i-1}$ and $g_i$ enclose $\CA_i$ for $1\leq i\leq m$;
\item $g_m$ and $g^*$ enclose $\CA^*$;
\item $X$ is adjacent to the struts of $R_i$ for $1\leq i\leq m$;
\item $Y$ is adjacent to the struts of $S_i$ for $1\leq i\leq m$.
\end{itemize}

\begin{lemma}
\label{lem:npredux}
Let $\II$ be an instance of numerical matching with target sums.
$\II$ has a solution if and only if
$G(\II)$ has a rectangular layout of width no more than $2\bmax+6$
and height no more than $6m+9$
(or vice versa).
\end{lemma}
\begin{proof}
The subgraph of $G(\II)$ induced by $X$, $Y$,
$t$, $b$, $g^*$, the $g_i$'s,
and paths through the accordions connecting
$t$, $g_0$, etc.~through $b$,
create a global $(m+4)$-rung ladder.
Scaling if necessary, we can assume all $\Delta_i\geq 3$.
Lemmas \ref{lem:ladder} and \ref{lem:enclosedaccord}
together imply that any layout
for $G(\II)$
respects the following (up to width/height symmetry).
(Refer to Figure \ref{fig:nprigid}.)
\begin{enumerate}
\item \label{np:i}
	The overall height is at least $6m+9$,
	because the $m+1$ enclosed accordions,
	each accordion itself of height 5, are all aligned,
	and the $m+4$ rungs of the global ladder are stacked.
\item \label{np:ii}
	The overall width
	is at least $2\bmax+6$:
	$2\bmax+2$ for $\CA^*$, $2$ to separate it from $X$ and $Y$,
	and $1$ each for $X$ and $Y$.
\end{enumerate}
(\ref{np:i}) is true regardless of whether
rectangle $t$ (rsp., $b$)
is above (rsp., below) $X$ and $Y$
or in between.
	\begin{figure}[t]%
	\begin{center}
	\input{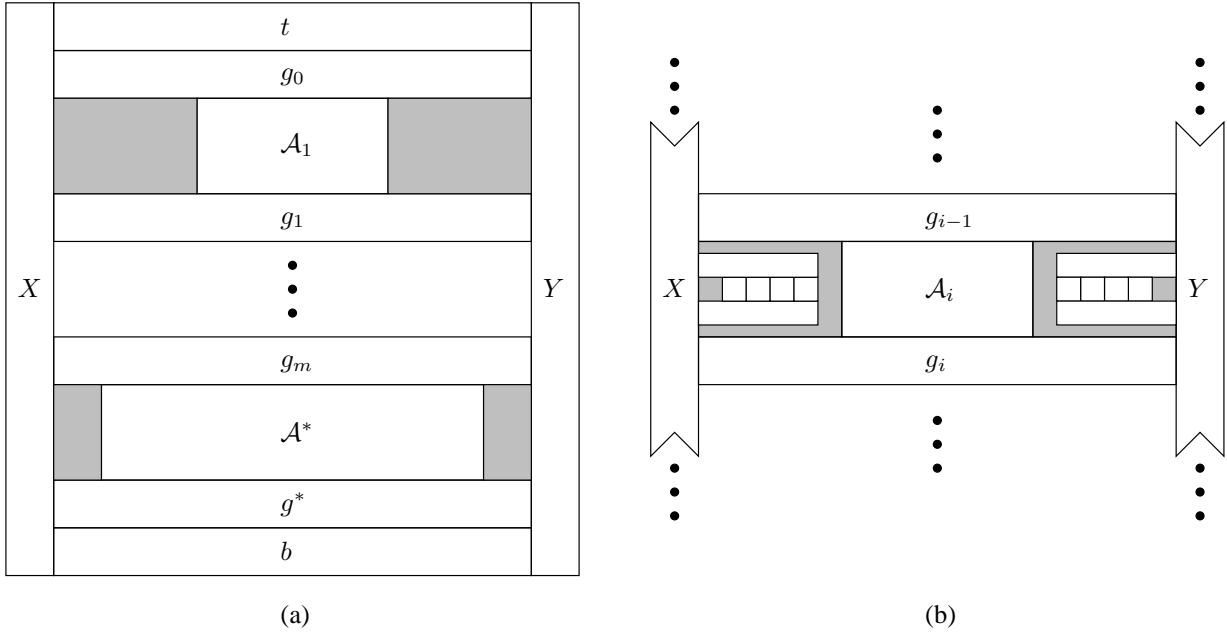}
	\end{center}
	\caption{
	(a) An incompressible sub-layout for $G(\II)$.
	(b) Filling the gaps in layer $i$ by some $R_i$ and $S_i$.
}%
	\label{fig:nprigid}%
	\end{figure}%

That is, the subgraph induces an {\em incompressible} sub-layout:
these components must have the indicated heights and/or widths
in any layout.

Refer to the sub-layout including $g_{i-1}$, $\CA_i$, and
$g_i$ as {\em layer} $i$.
Accordion $\CA_i$ creates gap(s) of total width $w(i) = 2\bmax+4 - \Delta_i$
between $X$ and $Y$ at layer $i$.
The idea is to fill each such gap with a sub-layout for
some ladder adjacent to $X$ or $Y$.
Each such ladder sub-layout must consist of two struts,
attached to $X$ (or $Y$),
with the rungs between them (except for possibly one of them) but separated
from $X$ and $Y$.
See Figure \ref{fig:nprigid}(b).
Ladder $R_i$ has $x_i$ rungs, and 
its sub-layout must therefore have length
at least $x_i+1$ ($x_i$ for rungs
plus one unit to separate the rungs from $X$).
Similarly, ladder $S_i$'s sub-layout must have length
at least $y_i+1$.

Assume \II~has a solution.
Re-index the elements of $X$ and $Y$
so that $x_i+y_i = B_i$ for $1\leq i\leq m$.
Then ladders $R_i$ and $S_i$ fit in the layer-$i$ gaps
without increasing the width
of each layer between $X$ and $Y$ from the minimum $2\bmax+4$.
To see this, note that
the width required for the sub-layouts
at layer $i$
is $x_i + y_i + \Delta_i + 4 = 2\bmax + 4$:
$x_i+1$ for $R_i$,
$y_i+1$ for $S_i$,
$\Delta_i$ for $\CA_i$,
and 2 units to separate $\CA_i$
from $R_i$ and $S_i$.
Also, each ladder sub-layout fits within the
height-5 lower bound 
of layer $i$.
Therefore, the width remains $2\bmax+6$,
the height remains $6m+9$,
and so the desired layout exists.

Assume $G(\II)$ has a layout of the hypothesized dimensions.
The incompressibility argument above implies that
each ladder must fit into a gap in one of the layers.
No two ladders may be placed in the same gap,
for the height of the layer would have to grow,
violating the height assumption.
Re-index the elements of $X$ and $Y$ so that
for $1\leq i\leq m$, $x_i$
and $y_i$ are the elements whose ladders, $R_i$ and $S_i$, fit into layer $i$.
The gap-height constraint implies that the bounding boxes for
$R_i$ and $S_i$
are neither above or below each other.
Their widths are thus bounded by the same
horizontal level of rectangles in $\CA_i$.
Therefore,
we can assume without loss of generality that $\CA_i$ is layed out
as a dissection
and thus by Lemma \ref{lem:enclosedaccord} in width $\Delta_i$.
Because $R_i$ and $S_i$ fit into the layer-$i$ gaps, 
it follows that $x_i+y_i+4 \leq w(i) = 2\bmax+4-\Delta_i$
and hence $x_i + y_i \leq B_i$.
As this is true for all layers,
Equation (\ref{eq:iiass})
implies that $x_i + y_i=B_i$ for all $i$,
which gives a solution to \II.
\end{proof}

\begin{theorem}
\label{thm:nphard}
Given a graph $G$ and values $W, H, A\in{\mathbb Z}^+$,
determining if $G$ has a (strong or weak)
rectangular layout of 
(1) width no more than $W$ and height no more than $H$
or (2) area no more than $A$
is NP-complete.
\end{theorem}
\begin{proof}
Theorem \ref{thm:grapharea} implies that the problem is in NP.
Lemma \ref{lem:npredux} provides a P-time
reduction showing NP-completeness,
because the number of vertices and edges in $G(\II)$
is $\poly(m)$
and numerical matching with target sums is strongly NP-complete:
for problem (1), set $W=2\bmax+6$ and $H=6m+9$,
and for problem (2), set $A=(6m+9)(2\bmax+6)$.
A similar reduction using height-3 accordions
can be used
for weak layouts.
\end{proof}

\begin{corollary}
\label{cor:nphard1dim}
Given a graph $G$ and value $L\in {\mathbb Z}^+$,
it is NP-hard to determine
the minimum width (rsp., height) layout
of $G$ such that the height (rsp., width) does not exceed $L$.
\end{corollary}

\subsection{Area Monotonicity}

We now explore differences between rectangular layouts and duals.
First we demonstrate that weak layouts, strong layouts, and duals
all have distinct area monotonicity properties.
Then we show that for graphs admitting both layouts and duals,
the different representations might require significantly
different areas.

For any graph $G=(V,E)$,
$V'\subset V$, and $E'\subset E$,
define $G_{V'}$ to be the induced subgraph on $V\setminus V'$
and $G_{E'}$ that on $E\setminus E'$ (removing isolated vertices).
By Corollary~\ref{cor:rlnft}, we know that both $G_{V'}$
and $G_{E'}$ have layouts if $G$ has a layout. 
The same does not necessarily hold for rectangular duals, however.
In general, given a {\em rendering strategy}---in this case
rectangular layouts or rectangular duals---we
say that a vertex or edge subset
is {\em rendering preserving} if the corresponding
subgraph defined above admits such a rendering.

Consider monotonicity of areas under augmentation.
For a given rendering strategy such that $A^*(G)$
is the area of an optimal rendering of $G$
(assuming $G$ admits a rendering),
we say the rendering is
{\em vertex} (rsp., {\em edge}) {\em monotone}
if for any graph $G$ 
and any rendering-preserving vertex subset $V'$ (rsp., edge subset $E'$)
it is true that
$A^*(G_{V'}) \le A^*(G)$
(rsp., $A^*(G_{E'}) \le A^*(G)$).

\begin{figure}[t]%
	\begin{center}
	\input{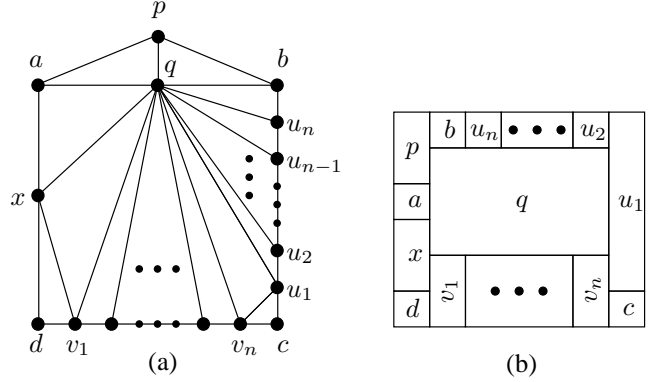}
	\end{center}
	\caption{(a) A graph on $2n+7$ vertices.
	(b) A rectangular dual of area $6(n+2)$.}%
	\label{fig:dmonot}%
	\end{figure}

\begin{figure}[t]%
	\begin{center}
	\input{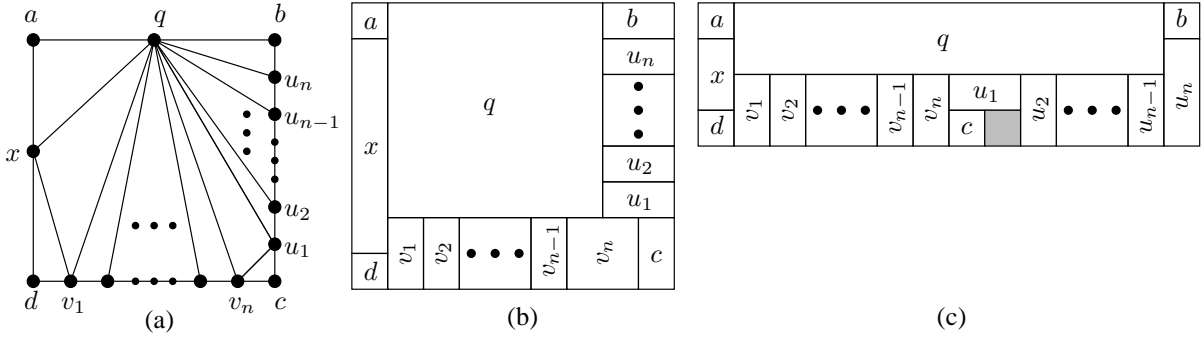}
	\end{center}
	\caption{(a) The graph on $2n+6$ vertices
		derived by deleting vertex $p$ from 
		Figure \ref{fig:dmonot}(a).
	(b) A rectangular dual of area $(n+3)(n+2)$.
	(c) A strong rectangular layout of area $4(n+2)$.
}%
	\label{fig:sldgap}%
	\end{figure}

\begin{theorem}
\label{thm:monot}
\hspace{1in}\newline\vspace{-3ex}
\begin{enumerate}
\item \label{mon:i}
	Weak layouts are vertex and edge monotone.
\item \label{mon:ii}
	Strong layouts are vertex monotone
	but not edge monotone.
\item \label{mon:iii}
	Rectangular duals are neither
	vertex nor edge monotone.
\end{enumerate}
\end{theorem}
\begin{proof}
(\ref{mon:i})
A weak layout for $G$ is also a weak layout for any subgraph of $G$.

(\ref{mon:ii})
Given a strong layout $L$ of $G$,
removing the rectangle corresponding to $v$
yields a strong layout of $G_{\{v\}}$ of no greater area;
hence, strong layouts are vertex monotone.
Figure \ref{fig:example}
disproves edge monotonicity, however:
by inspection, any strong layout for the graph
in Figure \ref{fig:example}(a) must have area at least $9$,
whereas the edge-augmented graph in Figure \ref{fig:example}(b)
has an area-6 strong layout.

(\ref{mon:iii})
For any $n\in{\mathbb Z}^+$,
Figure \ref{fig:dmonot} shows a graph $G$ on $2n+7$ vertices
and a rectangular dual of area $6(n+2)$.
Suppose, however, we delete vertex $p$.
In any rectangular dual of $G_{\{p\}}$
(as depicted in Figure \ref{fig:sldgap}(a)),
rectangles $a$, $b$, $c$, and $d$ must be the corners
of the dissection, for each has degree only $2$.
The height and width must therefore each be $\Omega(n)$,
as exemplified in Figure \ref{fig:sldgap}(b).
The area is thus $\Omega(n^2)$,
which disproves vertex monotonicity
and also edge monotonicity,
because the latter implies the former.
\end{proof}

\subsection{Gaps between Layouts and Duals}

Consider the potential gap between
minimum-area layouts and rectangular duals
of the same graph.
Define $A^*_s(G)$ (rsp., $A^*_D(G)$)
to be the area of an optimal strong rectangular layout
(rsp., rectangular dual) of $G$.

\begin{theorem}
\label{thm:gap}
There exists an infinite family $\CG$ of graphs
such that for any $G\in \CG$,
$A^*_s(G) = O(n)$ but $A^*_D(G)= \Omega(n^2)$. 
\end{theorem}
\begin{proof}
Define $\CG$ to contain the graph shown in Figure \ref{fig:sldgap}(a)
for each $n\in{\mathbb Z}^+$.
Consider some $n\in{\mathbb Z}^+$
and the corresponding $G\in\CG$.
As argued in the proof of Theorem \ref{thm:monot},
any rectangular dual of $G$ has area $\Omega(n^2)$,
while
Figure \ref{fig:sldgap}(c) depicts a strong rectangular layout
of height $4$ and width $n+2$.
\end{proof}

\section{Conclusion}
\label{sec:conc}

We have presented a new characterization of contact graphs of isothetic
rectangles in terms of those planar graphs that can be embedded with no
filled three-cycles.  
We have shown the general area and constrained-width
(and -height) optimization problems for rectangular layouts to be NP-hard
and provided $O(n)$-time algorithms to construct $O(n^2)$-area rectangular
layouts for graphs and $O(n\log n)$-area rectangular layouts for trees.

Many open problems remain.  
What is the
hardness of approximating the minimum-area rectangular layout?  Are there
better approximation algorithms than the ones we presented here
($O(n)$-approximation for graphs; $O(\log n)$ for trees)?  
Is approximating the minimum dimension (width or height)
easier than approximating the area?
This problem is motivated
by applications on
fixed-width, scrollable displays.
Also, does the NP-hardness result extend to rectangular
duals?  Since graphs that admit such duals are internally triangulated
(triangulated except for the outer face), the freedom to place components
to satisfy partition-type reductions does not seem to exist.

Can our techniques be applied to study contact graphs on other closed
shapes: for example, squares, arbitrary regular polygons, arbitrary convex
polygons, and higher-dimensional shapes?  Also, allowing corner touching
to imply adjacency changes the class of graphs described by layouts.  For
example, $K_4$ can be expressed by four rectangles meeting at a corner,
and $K_6$, which is not planar, can be expressed by triangles.  For $k\geq
4$, however, no layout on $k$-gons can express a non-planar graph.
Allowing corner touching also opens the question of allowing non-isothetic
rectangles, which can represent embeddings with filled triangles.
Finally, is there a class of polygonal shapes other than disks whose
contact graphs are the planar graphs?

%%% Local Variables: 
%%% mode: latex
%%% TeX-master: "main"
%%% End: 

\section*{Acknowledgements}
%\paragraph{Acknowledgements.}

We thank Therese Biedl 
and
Chandra Chekuri
for fruitful discussions
and
Anne Rogers for her tolerance.
We thank the anonymous referees for many constructive comments.

\bibliographystyle{plain}
\bibliography{compgeom,complexity,connect,facts,graph,graphdraw,selfadj,vlsi}

\newpage
\appendix

\section{Proofs of Lemmas}
\label{app:proofs}

\noindent
{\bf Lemma \ref{lem:ptc4ec}}~{\em \lemptc}\\[1ex]
\begin{proof}
$G$ is a planar triangulation if and only if $G^*$
is planar and cubic.
It therefore suffices to prove that
(1) if $G$ is a planar triangulation that is not 4-connected,
        then $G^*$ is not cyclically 4-edge connected,
and
(2) if $G^*$ is cubic and planar but not cyclically 4-edge connected,
        then $G$ is not 4-connected.

Assume that $G$ is a planar triangulation that is not 4-connected.
By Lemma \ref{lem:4ctri} $G$ has some separating triangle
$(a,b,c)$.
In $G^*$, therefore,
there are edges $\overline{\{a,b\}}$,
$\overline{\{b,c\}}$,
and $\overline{\{c,a\}}$,
each connecting
the two faces in $G$
incident upon edges $\{a,b\}$,
$\{b,c\}$, and $\{c,a\}$, resp.
These three edges separate $G^*$ into two components, $G^*_1$ and $G^*_2$.
That $(a,b,c)$ is a separating triangle
and $G$ is triangulated
implies that $G^*_1$ and $G^*_2$ contain cycles.
Hence $G^*$ is not cyclically 4-edge connected.

Assume that $G^*$ is cubic and planar but not cyclically 4-edge connected.
Then there exist three edges $e_1$, $e_2$, and $e_3$ that
separate $G^*$ into two or three components, each of which contains a cycle.
Assume for now a separation into two components,
$G^*_1$ and $G^*_2$.
Without loss of generality, each of $e_1$, $e_2$, and $e_3$
has an endpoint in each component.
Therefore there are vertices $v_1$, $v_2$, and $v_3$ in $G$
that correspond to the three faces induced in $G^*$
between pairs of $e_1$, $e_2$, and $e_3$.
Furthermore, $(v_1,v_2,v_3)$ forms a separating triangle,
because the cyclic nature of $G^*_1$ and $G^*_2$
implies that the corresponding components $G_1$ and $G_2$
in $G$ are non-empty.
By Lemma \ref{lem:4ctri}, therefore, $G$,
which is a planar triangulation, is not 4-connected.
A similar argument holds when $G^*$ is separated into three components,
in which case some pair among $e_1$, $e_2$, and $e_3$
separates $G^*$ into two components.
\end{proof}

\bigskip

\noindent
{\bf Lemma \ref{lem:cc4e}}~{\em \lemcc}\\[1ex]
\begin{proof}
Assume $G$ is a saturated plane map with no 3-ring.
By definition, $G$ is cubic and planar,
and so $G^*$ is a planar triangulation.
Furthermore,
because
a 3-ring in $G$ induces a separating triangle in $G^*$
and vice-versa,
it follows that
$G^*$ has no separating triangle.
By Lemma \ref{lem:4ctri},
therefore,
$G^*$ is 4-connected,
and by Lemma \ref{lem:ptc4ec},
$G$ is cubic, cyclically 4-edge connected.

Assume $G$ is planar, cubic, and cyclically 4-edge connected.
Then by Lemma \ref{lem:ptc4ec}
$G^*$ is a 4-connected planar triangulation.
By Lemma \ref{lem:4ctri},
$G^*$ has no separating triangle,
and so $G$ has no 3-ring.
That $G$ is a saturated plane map follows the assumption by definition.
\end{proof}

%%% Local Variables: 
%%% mode: latex
%%% TeX-master: "main"
%%% End: 

\end{document}